\documentclass[12pt,preprint]{aastex}
\usepackage{ulem}
\usepackage{color}
\usepackage{subfigure}

\slugcomment{To appear in ApJ.}

\shorttitle{Carbon-Chains in Warm Environments}
\shortauthors{Hassel, Harada \& Herbst}

\begin{document}

\title{Carbon-Chain Species in Warm-up Models}

\author{George E. Hassel}
\affil{Department of Physics, The Ohio State University,
    Columbus, OH 43210}
\affil{Present address: Department of Physics and Astronomy, Siena College, Loudonville, NY 12211}
    \email{ghassel@siena.edu}
    \author{Nanase Harada}
\affil{Department of Physics, The Ohio State University,
    Columbus, OH 43210}
\and
\author{Eric Herbst}
\affil{Departments of Chemistry,  Astronomy,  and Physics, University of Virginia,
    Charlottesville VA 22904 USA}

\begin{abstract}
In previous warm-up chemical models of the low-mass star-forming region L1527, we investigated the evolution of carbon-chain unsaturated hydrocarbon species when the envelope temperature is slightly elevated to $T\approx 30$~K.  These models demonstrated that enhanced abundances of such species can be explained by gas-phase ion-molecule chemistry following the partial sublimation of methane from grain surfaces.  We also concluded that the abundances of  hydrocarbon radicals such as the C$_{\rm n}$H family should be further enhanced as the temperatures increase to higher values, but this conclusion stood in contrast with the lack of unambiguous detection of these species toward hot core and corino sources.  Meanwhile, observational surveys have identified C$_2$H, C$_4$H, CH$_3$CCH, and CH$_3$OH toward hot corinos (especially IRAS 16293-2422) as well as towards L1527, with lower abundances for the carbon chain radicals and higher abundances for the other two species toward the hot corinos.  In addition, the {\it Herschel Space Telescope} has detected the bare linear chain C$_3$ in 50 K material surrounding young high-mass stellar objects.  To understand these new results, we revisit previous warm-up models with an augmented gas-grain network that incorporated reactions from a gas-phase network that was constructed for use with increased temperature up to 800 K.  Some of the newly adopted reactions between carbon-chain species and abundant H$_2$ possess chemical activation energy barriers.  The revised model results now better reproduce the observed abundances of unsaturated carbon chains under hot-corino (100~K) conditions and make predictions for the abundances of bare carbon chains in the 50 K regions observed by Herschel HIFI.  

\end{abstract}

\section{Introduction}

The synthesis and evolution of carbon-chain species have once again become topics of interest, specifically pertaining to early stages of low-mass star formation in which the envelope temperature is warmer than the parent dark cloud temperature of 10 K.  Recent observations indicate that two specific envelope stages for low-mass protostars, hot and lukewarm corinos, show differences in their abundances of carbon chain species. 

 Hot corinos are relatively condensed ($n_{\rm H}\sim 10^{7}$ cm$^{-3}$) and hot ($T\gtrsim 100~$K) regions, and feature a molecular composition rich in hydrogen-rich organic species as well as oxygen-bearing species.  They can be considered  low-mass analogs to the hot cores that surround high-mass star-forming regions.  Some of the first hot corinos to be categorized were IRAS 16293-2422, NGC 1333 IRAS 4A and 4B \citep{bclwcccmpt04,jsv04,mea04,cea03}. 
As in hot cores, the complex composition is the combined result of grain surface and gas phase processes \citep{gh06}.

Observations of L1527, an envelope surrounding the transitional Class 0/I protostar IRAS 04368+2557 in Taurus, indicated that enhanced abundances of unsaturated (hydrogen-poor) hydrocarbons are likely related to a slightly elevated envelope temperature of $T \approx 30$~K via a chemistry labeled ``warm carbon-chain chemistry''  (WCCC) \citep{sea07}.  Although the chemical composition resembles that of a dark cloud core,  the composition of WCCC regions can also be attributed to a process beginning with grain surface chemistry.    The elevated temperature allows methane to evaporate from the ice mantles surrounding dust particles and act as a precursor for a carbon-chain rich ion-molecule chemistry.
The resulting chemical composition is a defining characteristic of the source, which we designated a ``lukewarm corino".  The idea of WCCC has proved somewhat controversial, however, and alternative explanations, such as standard gas-phase ion-molecule chemistry in a collapsing source, have been advanced \citep{charn10}.   A recent interferometric study provides  evidence for 
WCCC chemistry within the central protostar and normal ion-molecule chemistry in cold cloud remnants 

\citep{sea11}. It has not been firmly established if the lukewarm corino phase is one step on an evolutionary track leading to the hot corino phase, or if it represents a terminal stage.

Since the observation of L1527, additional observations have identified a similar kinetic temperature and similar enhancements for hydrocarbon chains towards the  source IRAS 15398-3359 in Lupus, alternatively identified as B228 \citep{sea09}, so that B228 is the second lukewarm corino to be studied.  Also, \citet{sea09} reported detections of some carbon chains, especially C$_{4}$H, toward a number of hot corinos at temperatures near 100 K, but with abundances significantly lower than those in the two lukewarm corino sources,  suggesting a diminution in abundance with increasing temperature.  More recently, linear C$_3$ has been detected in absorption in regions at a temperature of $\approx$ 50~K  against continua in the THz region from the sources W31C and W49N \citep[hereafter referred to as the HIFI sources]{mea10c3}.    

We previously reported the use of a gas-grain chemistry network with a simple warm-up following a cold stage to simulate the evolution of molecular abundances in L1527 as a result of increasing temperature \citep{hhg08}(hereafter HHG). A comparison of model abundances with the observations of \citet{jsv04} and \citet{sea08,sea07} showed that the best overall agreement occurs at times when the temperature approaches $T=30$ K, the indicated temperature for the envelope.  Reasonable agreement with observation in L1527 was found near this temperature whether or not the warm-up was terminated at 30 K or continued to higher temperatures.  Similar chemistry was also found in the hydrodynamic-chemical model of prestellar to protostellar core formation of \cite{aik08}.  

Some of the warm-up models were extended to maximum temperatures of 100 and 200 K, in a similar fashion to earlier hot core models, in which it was shown that the standard oxygen-containing complex molecules in such sources could be partially produced by surface chemistry followed by desorption \citep{gh06,gwh08}.  These prior studies emphasized the typical hot-core species instead of the calculated abundances of carbon-chain species.  It was later noted  that in all of our warm-up models, carbon chain species are quite abundant at the higher temperatures for long periods of time.  This result contrasts with the lack of strong detections of such species toward hot core and hot corino sources, with the general exception of HC$_{3}$N \citep{hvd09}.  In the survey of \citet{sea09}, the fractional abundance values obtained for C$_{4}$H vary, but are generally smaller than $X_{\rm C_{4}H} = N_{\rm C_{4}H}/N_{\rm H} =10^{-10}$ in hot corinos, as compared with $X_{\rm C_{4}H} \approx$ 2--3 $\times 10^{-9}$ for the two lukewarm corino sources.  The observed hot corino result is smaller than the peak fractional abundance of $X_{\rm C_{4}H} > 10^{-8}$ found in the extended warm-up models of HHG.

Some other observational studies have been directed at the question of unsaturated molecules in star-forming regions. In their review of the chemistry of complex molecules, \citet{hvd09} mentioned a study of the C$_{4}$H/CH$_{3}$OH and HC$_{5}$N/CH$_{3}$OH abundance ratios in seven hot cores that resulted in upper limits of 0.01--0.1 for the ratios (S. and J. J{\o}rgensen-Bisschop, private communication).  In addition, an average abundance of $N_{\rm CH_{3}OH}/ N_{\rm H} = 9.5\times 10^{-7}$ was reported for the seven hot cores \citep{BJDW07}, leading to upper limits for C$_{4}$H and HC$_{5}$N high enough to overlap with the $T_{\rm max} = 100$ \& 200 K peak results of HHG for both species.

It would appear at present that C$_{4}$H is detected in largest abundance in the two known lukewarm corinos, that it is detectable in hot corino sources although to much lower abundance than predicted, and that it has a high upper limit with respect to methanol in hot cores.  The survey of \citet{sea09b} also included the detection of CH$_3$OH toward L1527 with a fractional abundance of $X_{\rm CH_{3}OH}\approx 10^{-9}$, similar to what is observed in cold cores such as TMC-1.  Unlike C$_{4}$H, and possibly other radicals of this class, methanol is a molecule typically associated with high gas-phase abundances only in hot cores and corinos, where it is desorbed from grains.  Although the picture is clearly incomplete at the present time, comparison of carbon chain species and species known to prefer high temperatures  provides an initial basis for chemical distinction of lukewarm and hot corino sources.  Although one can think of a lukewarm corino as a prior stage of evolution than a hot corino,   \citet{sea09b} have also suggested that the difference between lukewarm and hot corinos has to do at least partially with the rate of collapse of the protostar; the slower the rate, the more chance for CO to form in the gas, accrete onto grains, and lead to oxygen-containing species; the faster the rate, the greater the chance that C atoms land on grains and form methane.    

In a recent study, \citet{har10} reported the extension of the temperature range of the OSU gas-phase chemical model  to 800 K by the inclusion of many new reactions, including those with non-zero activation barriers.  Select reactions might also be relevant at hot core and corino temperatures.  In particular, reactions between the carbon chain radicals C$_{\rm n}$H and C$_{\rm n}$ and H$_2$ were omitted from previous networks because they possess barriers of $E_{\rm{A}} \gtrsim 10^3$ K, and are inefficient in cold models.  The effect of adding these gas-phase reactions to our warm-up gas-grain model network would likely be to decrease the calculated abundances of these radicals at higher temperatures, so as to make the results more in line with hot corino observations of C$_{4}$H.  On the other hand, activation energy barriers for reactions between normal molecules in ground electronic singlet states, such as cyanopolyynes, and H$_{2}$ should be much larger so that these reactions are still not likely to be important in hot corinos.  The updated reaction network should not affect the abundances at lukewarm corino conditions.  In this paper, we present revised gas-grain warm-up models including addition of gas-phase radical-H$_{2}$ reactions as well as many other reactions added by \citet{har10}, and use the results to determine a revised temperature dependence of the abundances of carbon-chain species in the range from 30~K (lukewarm corinos) to 100~K (hot corinos).   

\section{Model}
\label{modelsection}

We utilize an updated version of the Ohio State University (OSU) gas-grain  code to follow the evolution of abundances during a warm-up to simulate the central regions surrounding a burgeoning protostellar core \citep{hhl92,gh06}.  In the model, a one-point parcel of material 

undergoes a gradual temperature increase following an initially cold period of $T_{0}=10$~K with a duration of $t=10^5$ yr.  This initially cold period allows for the accumulation of significant solid-phase abundances of H$_2$O(s), CO(s) and CH$_4$(s).  Varying the length of the cold period between 10$^{4}$-10$^{6}$ yr did not significantly change the results of HHG, so we presently consider models that begin warming up at 10$^{5}$~yr in the interest of making direct comparisons with prior results. 

We consider constant gas densities of $n_{\rm H} = 2\times 10^{5}$~and~$1 \times10^{6}$~cm$^{-3}$ based on estimates for the HIFI and lukewarm corino sources, respectively.  The same physical parameters and initial chemical abundances as found in  \citet{gwh07} and Tables 1 \& 2 of HHG are adopted.   Our approach uses homogeneous physical conditions at any given time, which is a simplification not found in more complex hydrodynamic approaches \citep{aik08}.  The advantage of this approach is that it allows a more detailed look at the chemical processes and the roles of individual reactions.

The warm-up rate is again treated according to
\begin{equation}
\label{eq-warmup}
T=T_{0}+(T_{\rm{max}}-T_{0})\left({{\Delta t} \over {t_{\rm{h}}}}\right)^{n},
\end{equation}
where we use values for the maximum temperature $T_{\rm{max}}$ of 30 K, 50 K, 100 K, and 200 K.  The 30 K value is used as a terminal temperature for lukewarm corino models, while the 100 K and 200 K values are used to represent sequences in which a lukewarm corino proceeds towards a hot corino.  The maximum temperature of 50 K  is used for HIFI models.

A heating timescale of $t_{\rm{h}}=2\times 10^5$~yr is adopted, with an order of $n=2$ following \citet{gh06}.  In addition to the warm-up models, we include a 10 K constant temperature model.  

The updated gas-grain network contains 7166 reactions involving a total of 668 gaseous and surface species.  Among the new gas-phase reactions, the most important are several sets of hydrocarbon radical-H$_{2}$ gas-phase reactions with moderate barriers, and reactions that distinguish the linear/carbene isomers HC$_3$ and H$_2$C$_3$ from the cyclic forms c-C$_3$H and c-C$_3$H$_2$.   The three new reaction classes that are most important for simulating radical chains in hot corinos are listed in Table~\ref{tbl-newrxns}.  The table also includes parameters for computing the modified Arrhenius rate coefficient of a gas phase reaction: 
\begin{equation}
k_{j} = \alpha {\left( {{T}\over{300~\rm{K}}}\right)}^{\beta} \exp{\left( {{-E_{\rm A}}\over{T}} \right)},
\end{equation}
where $E_{\rm A}$ (more commonly referred to as $\gamma$) represents the activation barrier.  The rate coefficients used for the first class (C$_{\rm n}$H + H$_{2}$) are based on assorted studies of the exothermic reaction with n = 2 \citep[for example]{far93,her94}.  Since the reported values for the rate coefficient show some dispersion, we have chosen to adopt the modified Arrhenius parameters from shock models \citep[G. Pineau des For\^{e}ts, private communication]{har10}, which should be reasonably accurate in the temperature range of interest.  Similarly, the rate coefficients used for the second class (C$_{\rm n }$H + H$_{2}$) are based on the value for n=2 from shock-model compilations \citep{PFD88}.  This reaction is also exothermic.  For the third set of reactions (C$_{\rm n}$N + H$_{2}$), we used the measured values for the exothermic CN + H$_{2}$ reaction \citep{UDFA07}.  The backwards endothermic reactions for these three reaction classes are ignored.
 
\section{Results}
\label{resultssection}

In this section, we report the effects of the expanded reaction network on various warm-up models for different classes of carbon-chain species.  We contrast the evolution of these species with that of methanol, a species more typically associated with hot corinos and hot cores.  We consider a broader comparison with observational results for corinos, both lukewarm and hot, in \S\ref{sec-obs}.    

\subsection{Polyyne Radicals: C$_{\rm n}$H}

\begin{figure}[b]
\centering
\subfigure{
\includegraphics[angle=90,scale=.3]{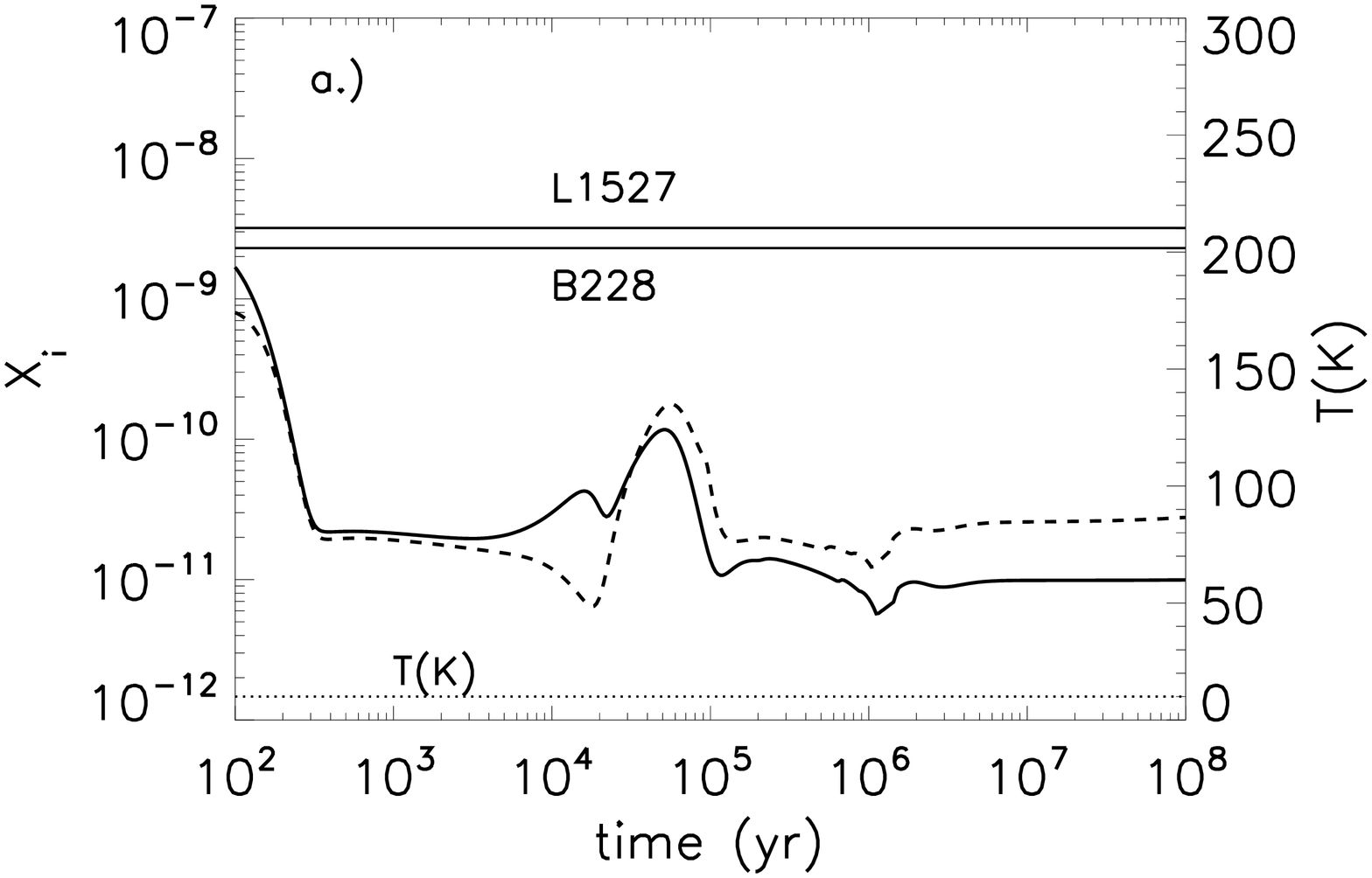}
}
\subfigure{
\includegraphics[angle=90,scale=.3]{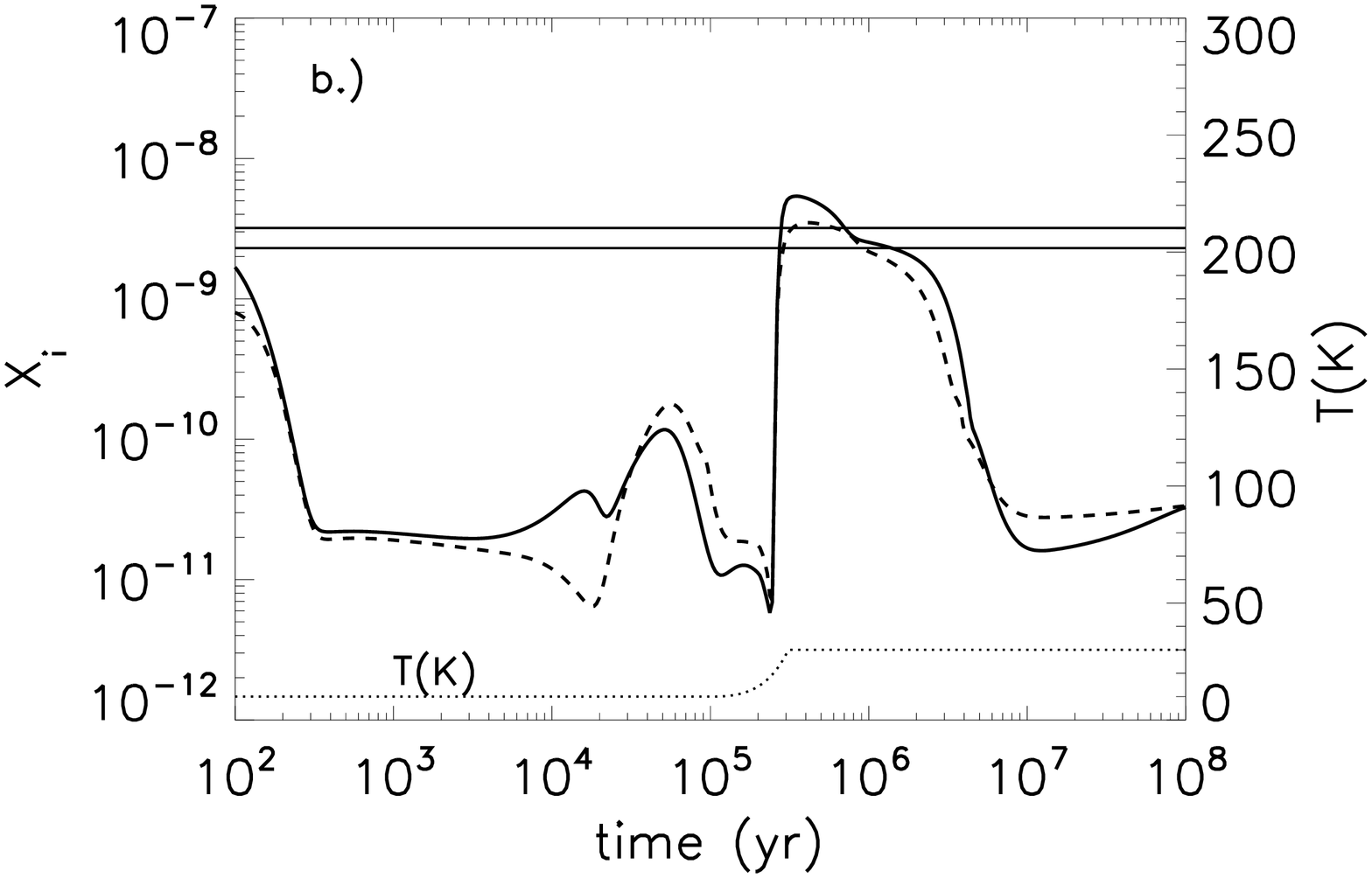}
}
\subfigure{
\includegraphics[angle=90,scale=.3]{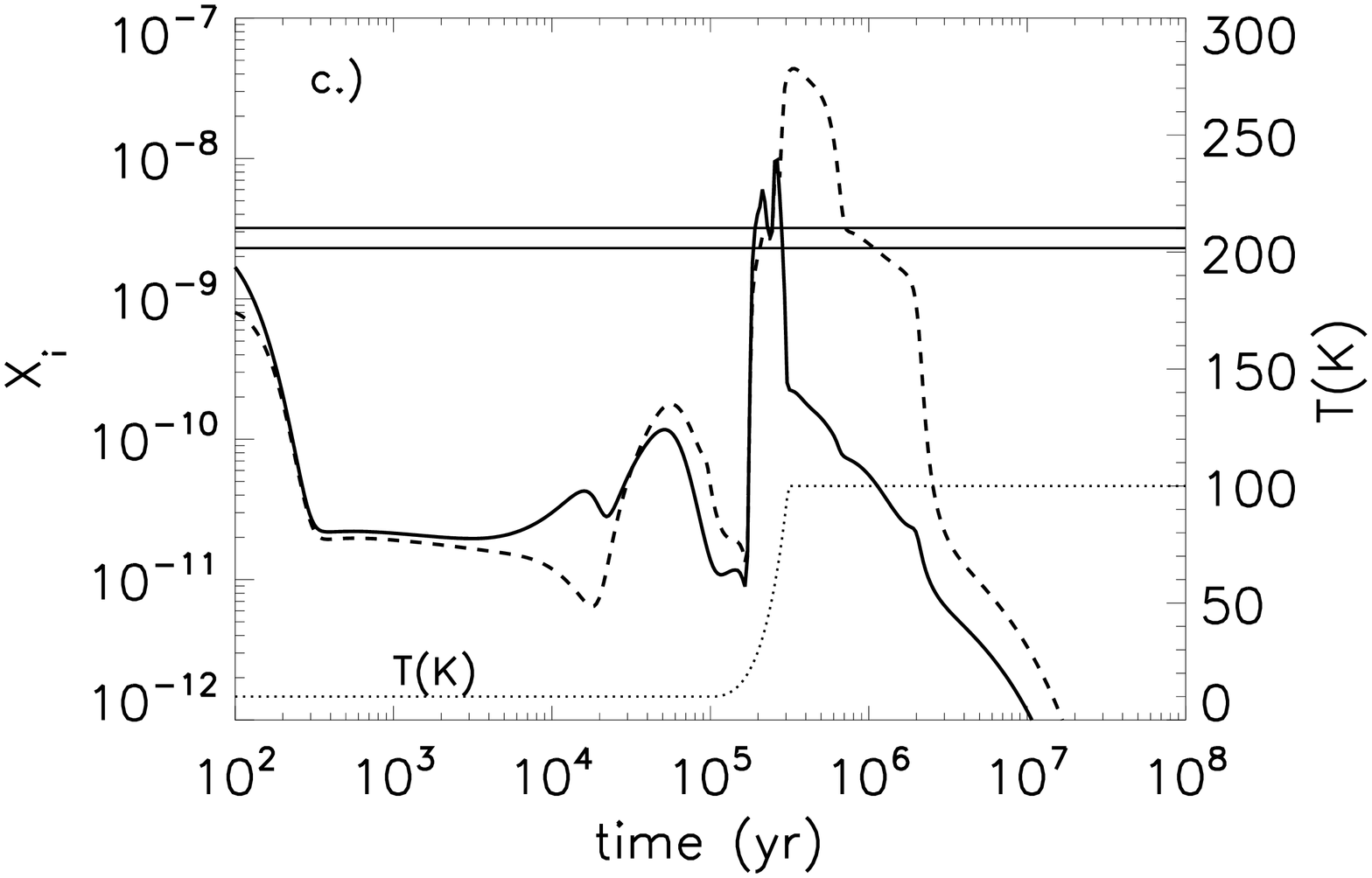}
}
\subfigure{
\includegraphics[angle=90,scale=.3]{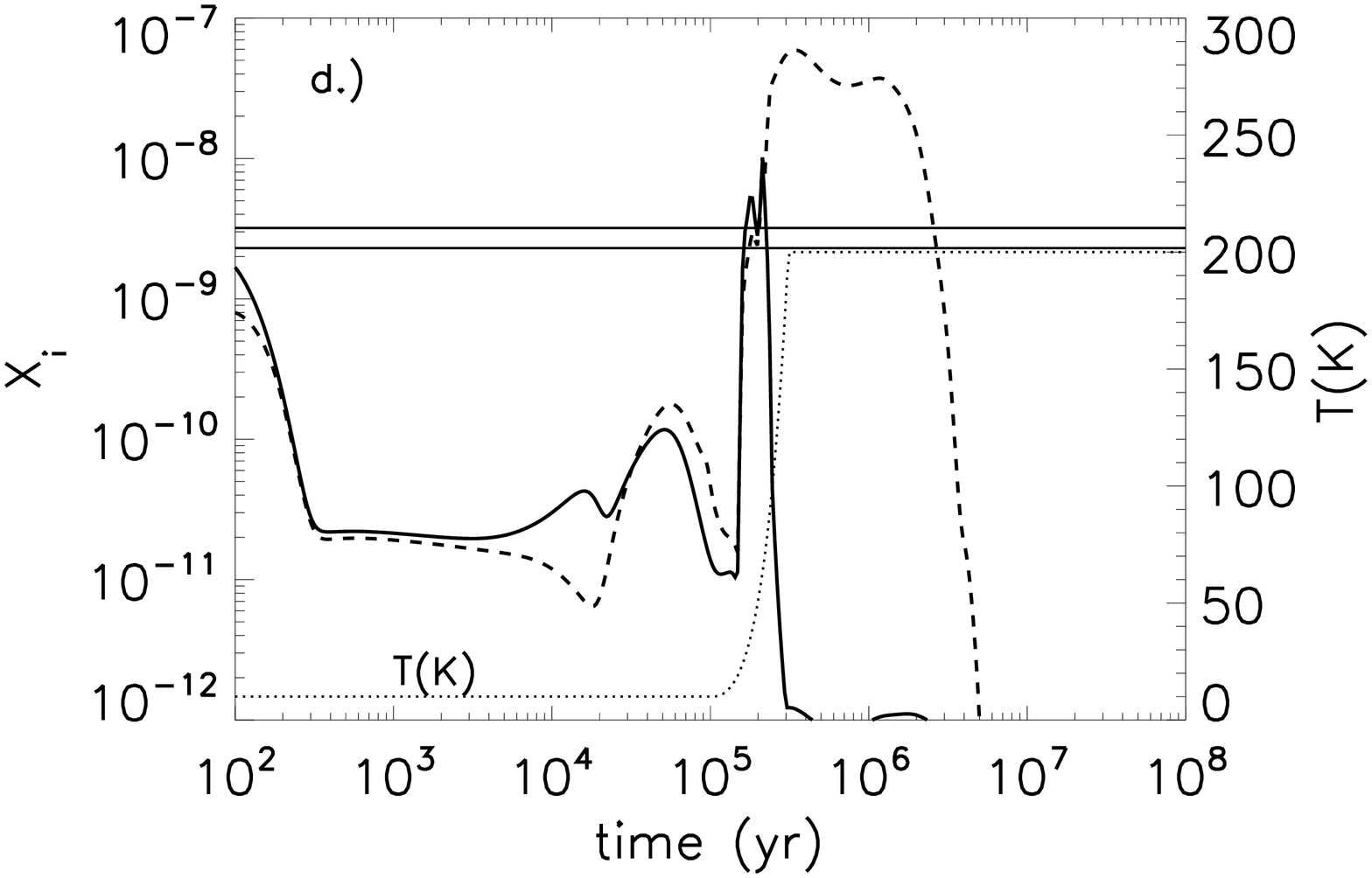}
}
\caption{Temporal evolution of the C$_4$H abundance for a variety of warm-up scenarios: (a) constant $T=10$ K; (b) $T_{\rm max} = 30$~K; (c) $T_{\rm max} = 100$~K; (d) $T_{\rm max} = 200$~K.  The solid lines represent new results including non-zero barrier reactions, while the dashed lines represent previous results from \citet{hhg08} for the same warm-up conditions.  The temperature profile for the warm-up is also plotted with respect to time as labeled.  The observed abundances in L1527 and B228 are indicated with horizontal lines.  \label{fig-C4H}}
\end{figure}

Figure~\ref{fig-C4H} shows the evolution of the polyyne radical C$_{4}$H in assorted models with the updated reaction network (solid lines) and the previous results  (dashed lines).  Panel a) contains our constant-temperature, cold (10 K) model, whereas the three other panels contain the results of warm-up models, in which the warm-up phase occurs from $1-3 \times 10^{5}$ yr, followed by a phase at the asymptotic temperature of 30~K, 100~K, and 200~K,  in panels b), c), and d) respectively.  The observational abundances for the two lukewarm corinos are shown as horizontal lines.  The results in panel a) show comparatively low abundances of C$_{4}$H  compared with observation for both old and revised models.   Panel b), in which the asymptotic temperature is 30 K, can be construed as a model of a lukewarm corino that remains at 30 K rather than continuing to higher temperatures.  Here one can see little change between the old and new results, indicating that up to 30 K, the new  reactions have little effect.  Panels c) and d) show our warm-up models leading to $T_{\rm max} = 100$~K and 200 K, roughly corresponding to the conditions of hot corinos and hot cores, respectively.  The C$_4$H abundances rise to values comparable with the peak of the $T_{\rm max}=30$~K model as the temperatures approach 30 K, in a similar manner to our prior model.   As the temperature continues to rise, however, a difference between the old and new models emerges.  In the older calculations, the C$_{4}$H abundance continues to rise to a peak value that exceeds 10$^{-8}$ and remains larger than $10^{-9}$ for a time period of  $t\approx 10^{6}$ yr, whereas in the updated ones, the  C$_{4}$H abundance almost immediately declines.

\begin{figure}[b]
\centering
\subfigure{
\includegraphics[angle=90,scale=.3]{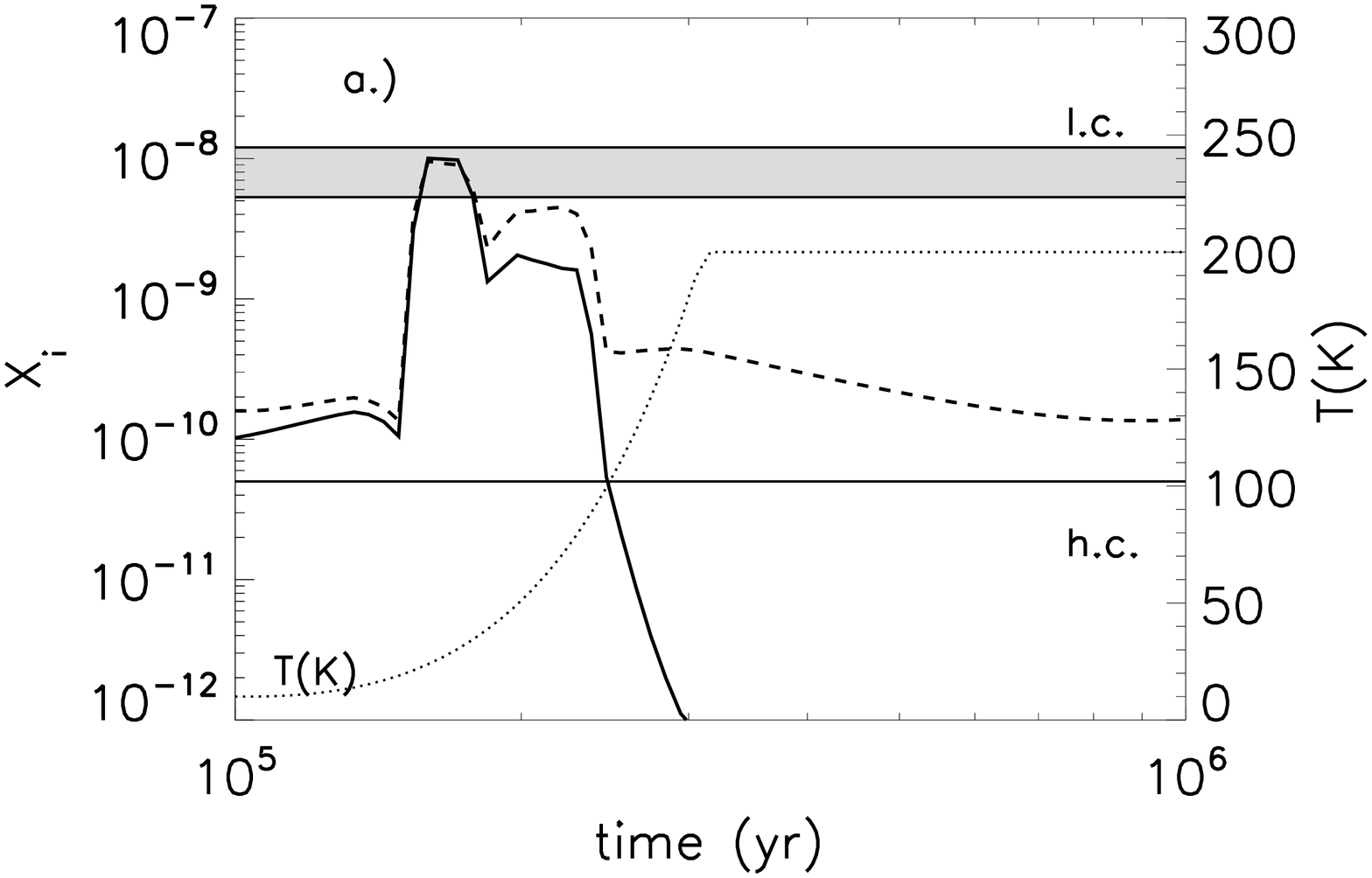}
}
\subfigure{
\includegraphics[angle=90,scale=.3]{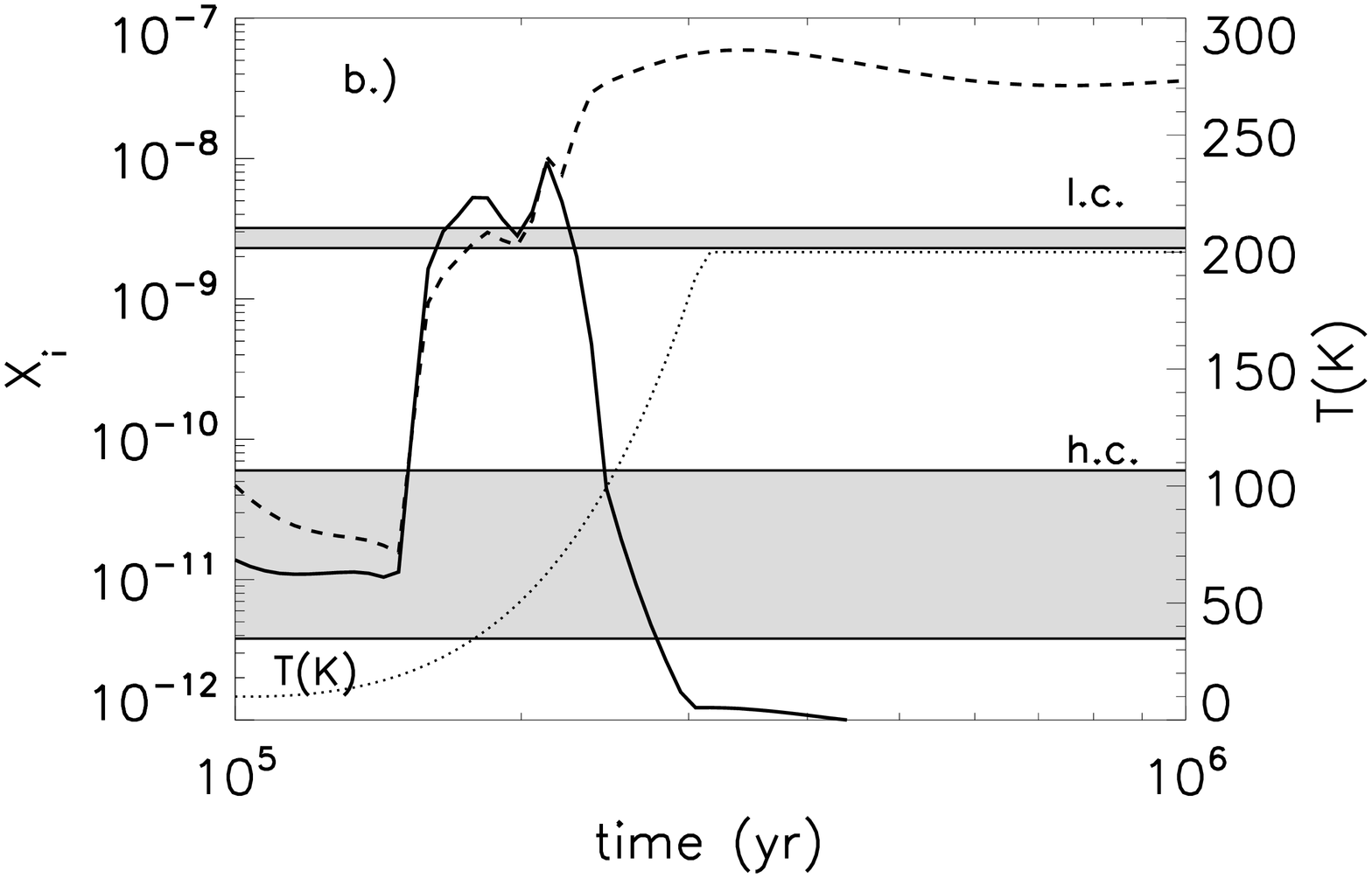}
}
\subfigure{
\includegraphics[angle=90,scale=.3]{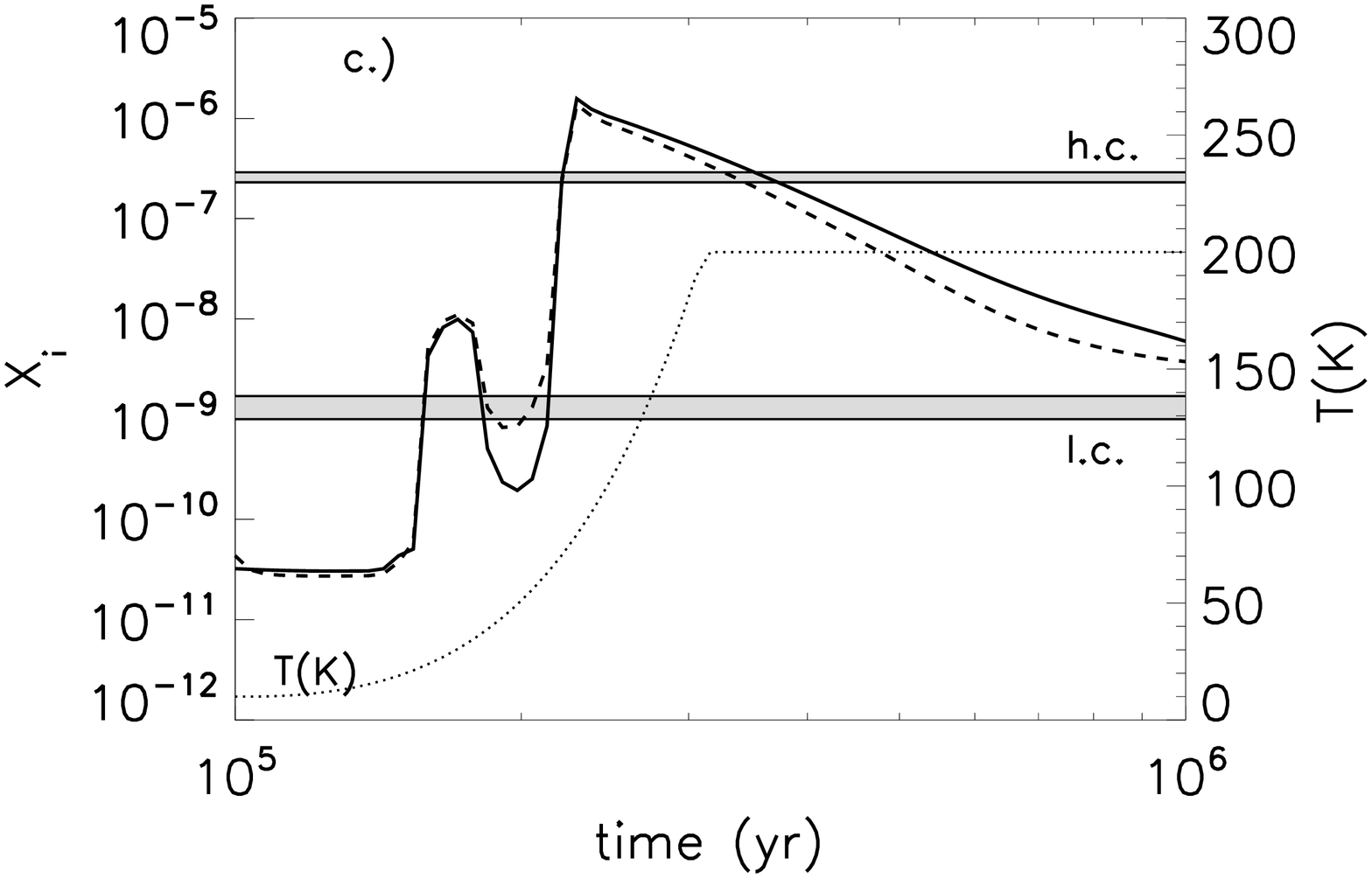}
}
\subfigure{
\includegraphics[angle=90,scale=.3]{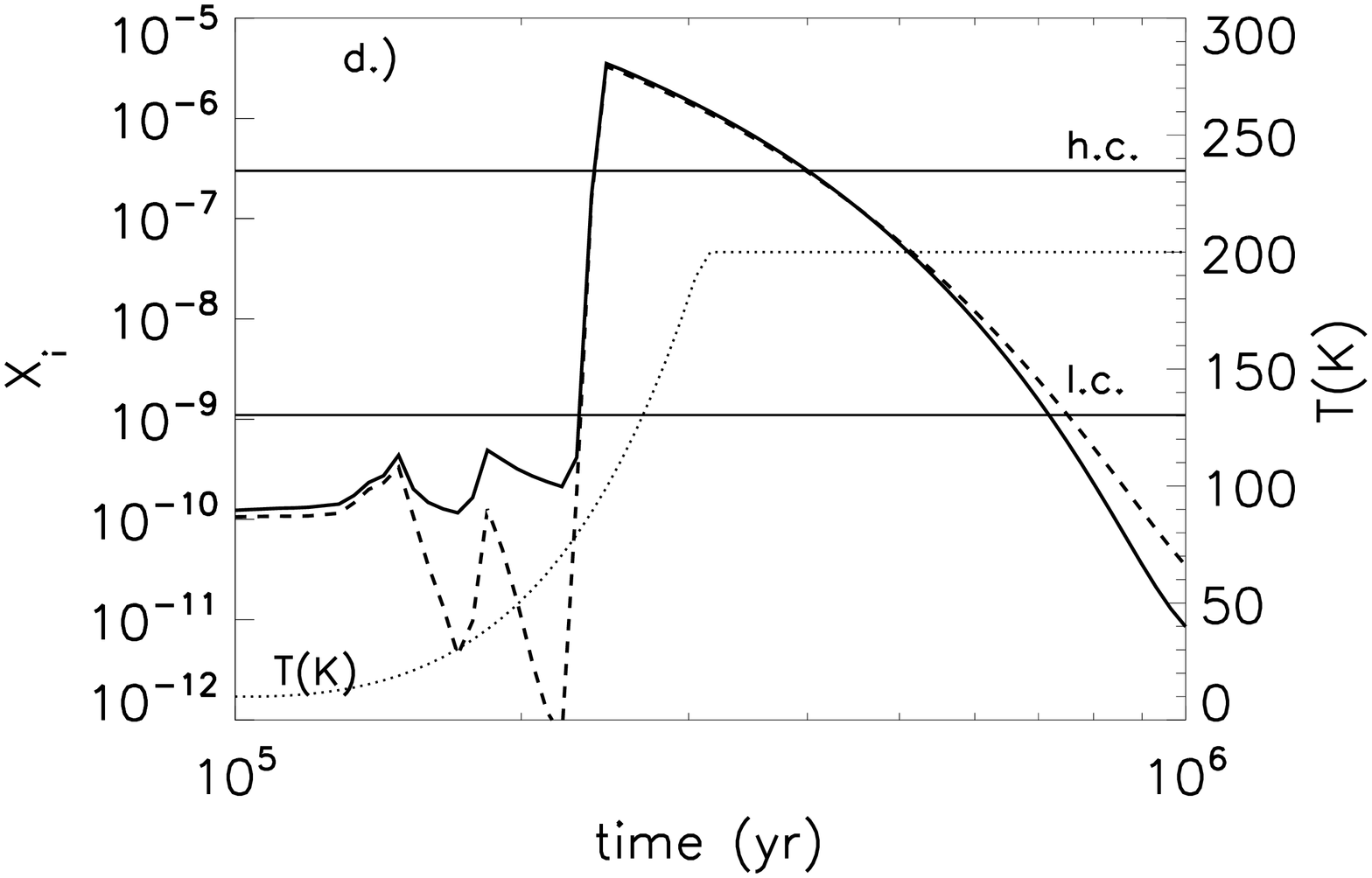}
}
\caption{Temporal evolution of the fractional abundances of four species: a.) C$_2$H; b.) C$_4$H; c.) CH$_3$CCH; d.) CH$_3$OH during the warm-up period to $T_{\rm max} = 200$~K.  The solid lines represent new results including non-zero barrier reactions, while the dashed lines represent previous results from \citet{hhg08}.  The temperature profile for the warm-up is also plotted with respect to time as labeled.  The observed abundances in the lukewarm corinos L1527 and B228 (as l.c.), and the hot corinos IRAS 16293-2422 and for some species NGC 133 IRAS 4B (as h.c.) are indicated with horizontal lines and grayscale where an observational range exists.    \label{fig-4spec}}
\end{figure}

Figure~\ref{fig-4spec} amplifies the abundance change of C$_{4}$H and other species (C$_{2}$H, CH$_{3}$CCH, CH$_{3}$OH) for the model in which the warm up continues all the way to 200 K. The  results for C$_{4}$H appear in panel (b), where some more details of the differences between the previous and newer models can be seen in the closer view.  After the temperature reaches $\approx 80$~K, the major distinction becomes apparent, as the reaction C$_4$H + H$_2$ $\rightarrow$ C$_4$H$_2$ + H becomes the dominant destruction mechanism in the newer model, and the abundance of C$_{4}$H declines precipitously.  

\begin{figure}[b]
\centering
\subfigure{
\includegraphics[angle=90,scale=.25]{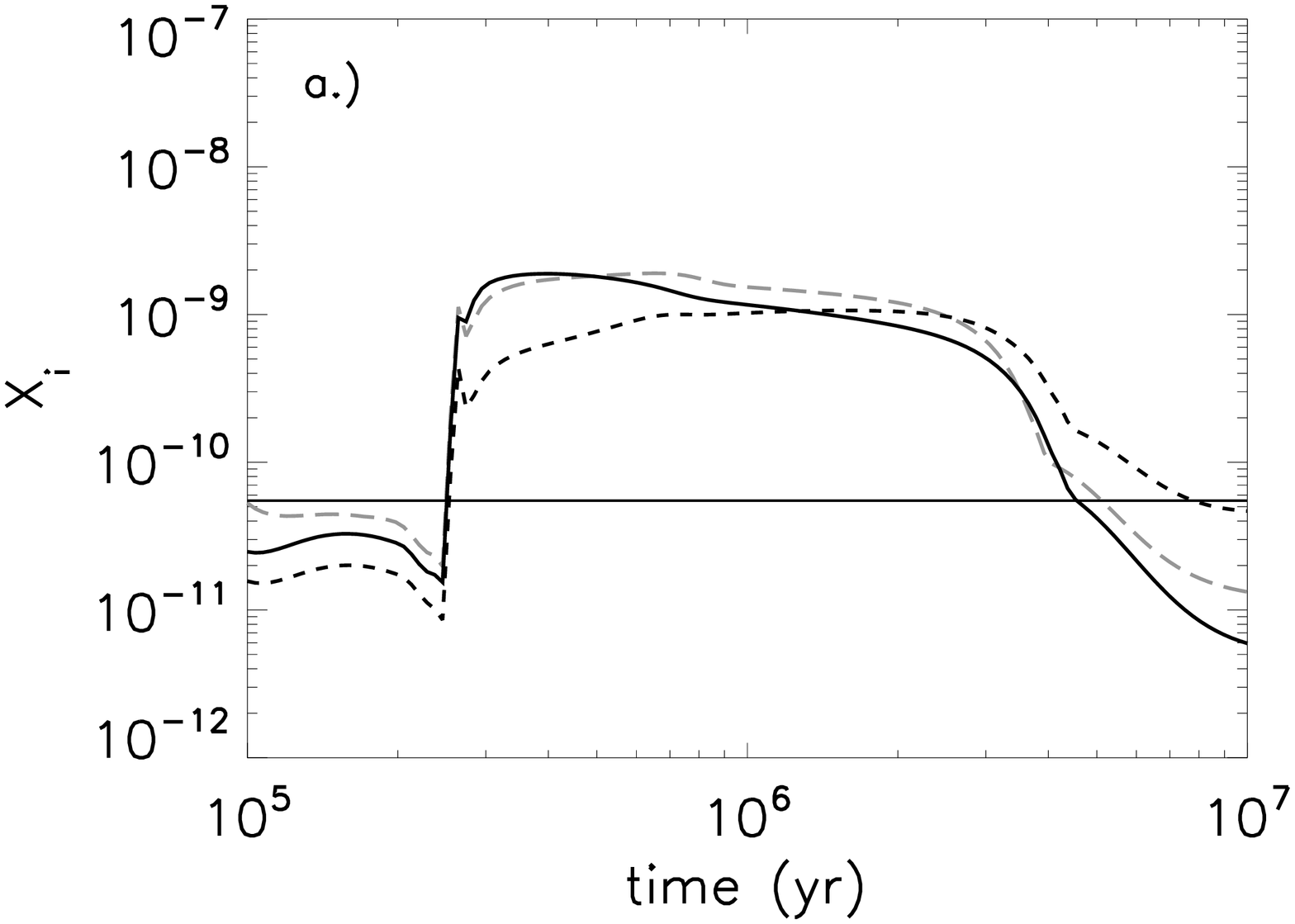}
}
\subfigure{
\includegraphics[angle=90,scale=.25]{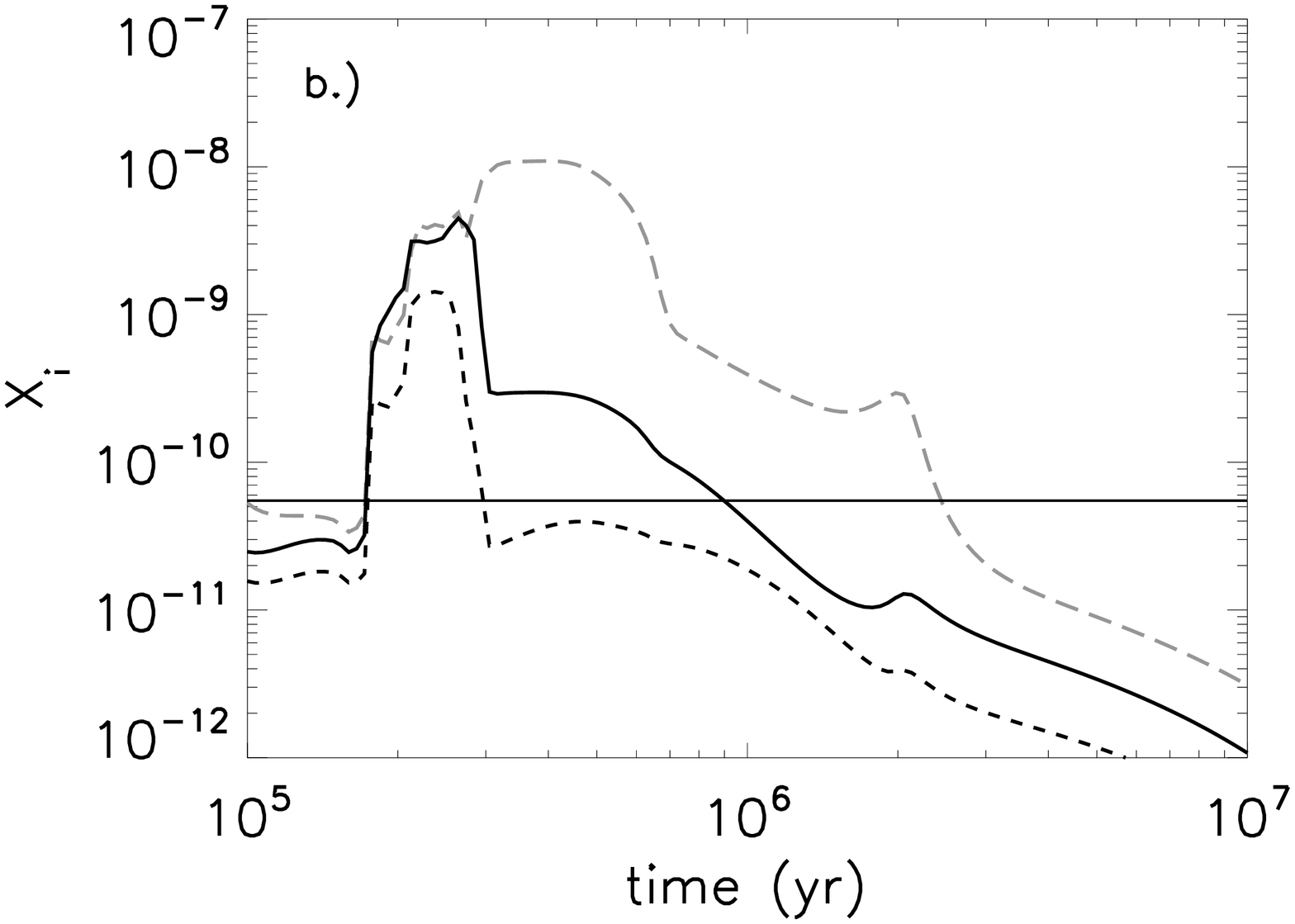}
}
\subfigure{
\includegraphics[angle=90,scale=.25]{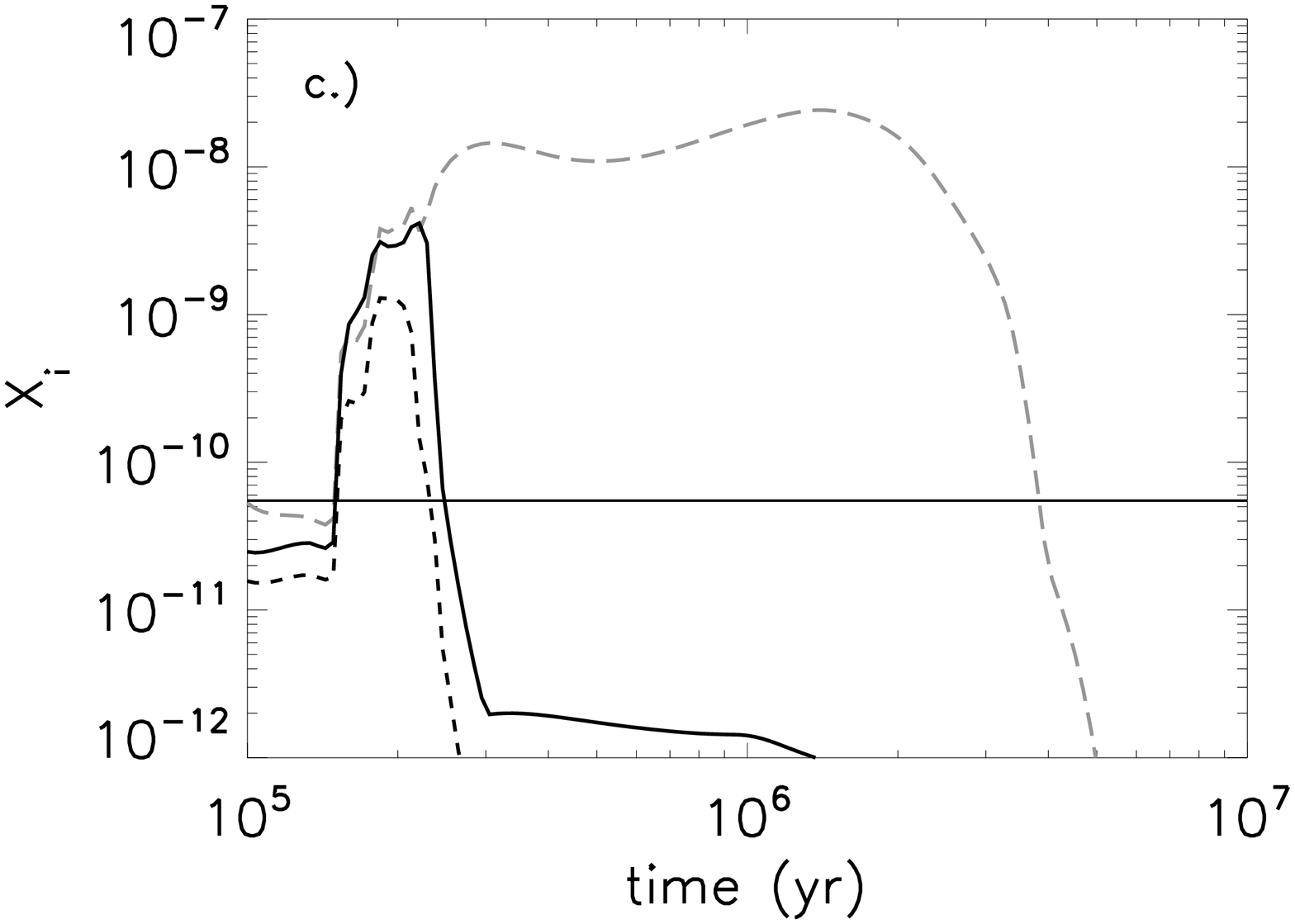}
}
\caption{Temporal evolution of the abundances of the cyclic and linear forms of C$_3$H for a variety of warm-up scenarios: $T_{\rm max} = $~(a) 30 K; (b) 100 K; and (c) 200 K.  The solid black lines represent the cyclic form, while the dashed black lines distinguish the linear HC$_3$ form.  Gray long-dashed lines represent the previous  results, in which no distinction was made between the isomers.  The horizontal line indicates the observed abundance of the linear isomer in L1527 \citep{sea09b}. 
 \label{fig-C3H}}
\end{figure}

The  abundances for the C$_{\rm n}$H chains evolve in a similar manner to C$_{4}$H in the assorted models.   The constant $T=10$~K model results agree between newer and previous models to the same degree as the results appearing in Figure~\ref{fig-C4H}a, and the $T_{\rm max}=30$~K warm-up model results rise to a similar peak value for a similar duration with bold old and new models.   Likewise, the $T_{\rm max}=100$~and 200 K model results rise to similar peak values as the corresponding 30 K results, but differences appear beyond this time in these models as a consequence of the inclusion of radical-H$_{2}$ reactions, which reduce the radical abundances quickly in the new calculations.   For the case of C$_{3}$H, there is a distinction in the new calculations between the linear (HC$_3$) and cyclic (c-C$_3$H) forms, with the latter having a somewhat higher abundance, as shown in Figure~\ref{fig-C3H}, which contains plots of the abundances vs time for three warm-up models.  As expected, we note a relatively decreased abundance at higher temperatures for both isomeric forms as compared with the previous results, as a result of the onset of high-temperature reaction with H$_2$.   

\subsection{Polyacetylenes, Rings, and Carbenes: HC$_{\rm n}$H, c-C$_{\rm n}$H$_{2}$, and H$_{2}$C$_{\rm n}$}

The products of the reactions in Table~\ref{tbl-newrxns}~imply that if these destruction reactions reduce the abundance of C$_{\rm n}$H chains at higher temperatures, there should be a corresponding increase in C$_{\rm n}$H$_2$ species.  This effect is seen at certain times for selected molecules, but the C$_{\rm n}$H$_2$ species can exist as different isomers, including the non-polar polyacetylene form HC$_{\rm n}$H, the carbene form H$_2$C$_{\rm n}$, and various ring forms  c-C$_{\rm n}$H$_2$.  This can make verification of models difficult both because reaction networks do not always distinguish the isomers and because only some of them are polar and have detectable rotational spectra.  In our previous analysis, we assumed that the carbene forms  represent 2$\%$ of the abundance of the undifferentiated model abundances, and we generally found agreement within an order of magnitude between observations and model results with this modification.  The assumption was based on the assumed rapid destruction with atomic oxygen for the carbene but not the more stable acetylenic species.   

\begin{figure}[b]
\centering
\subfigure{
\includegraphics[angle=90,scale=.25]{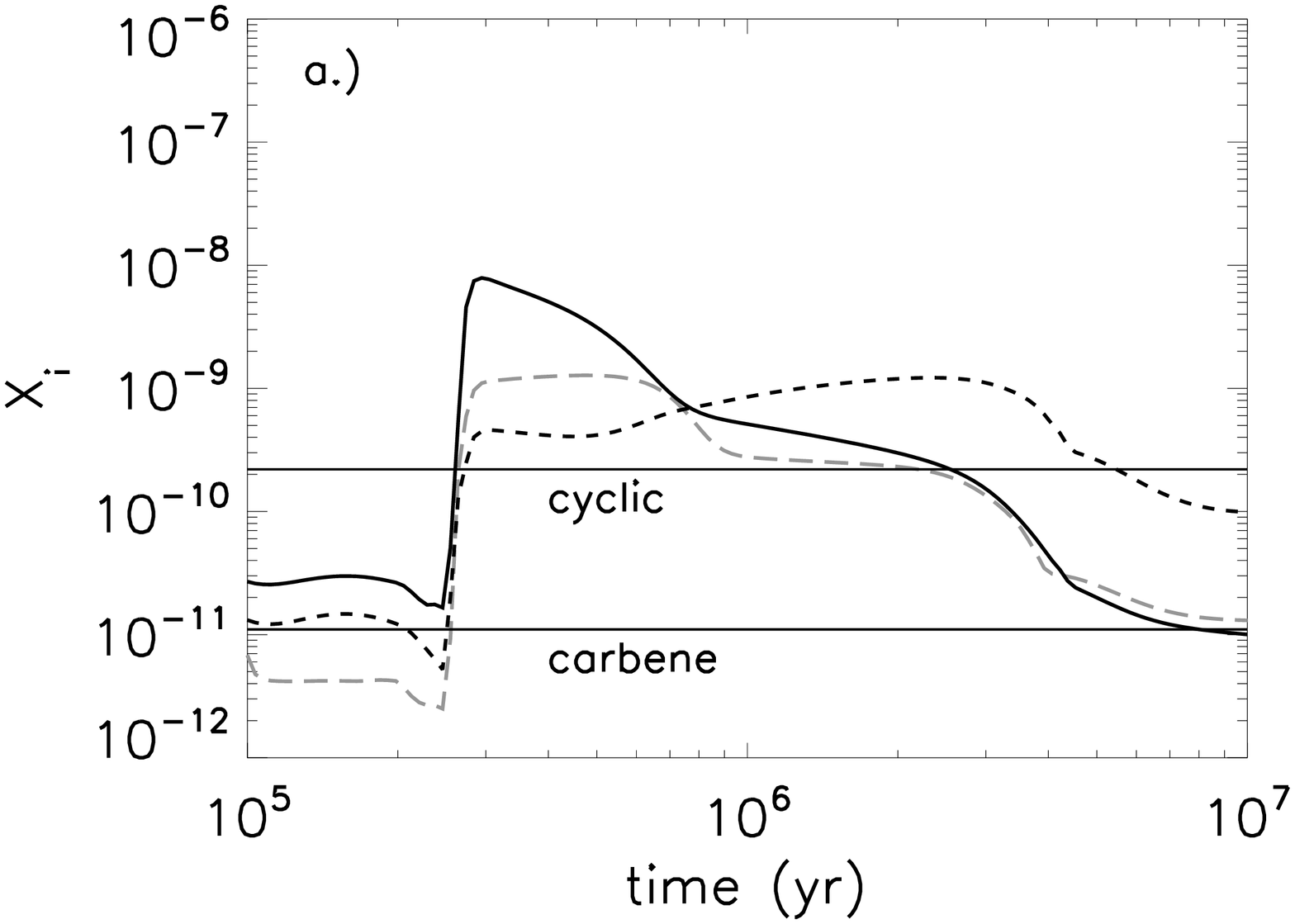}
}
\subfigure{
\includegraphics[angle=90,scale=.25]{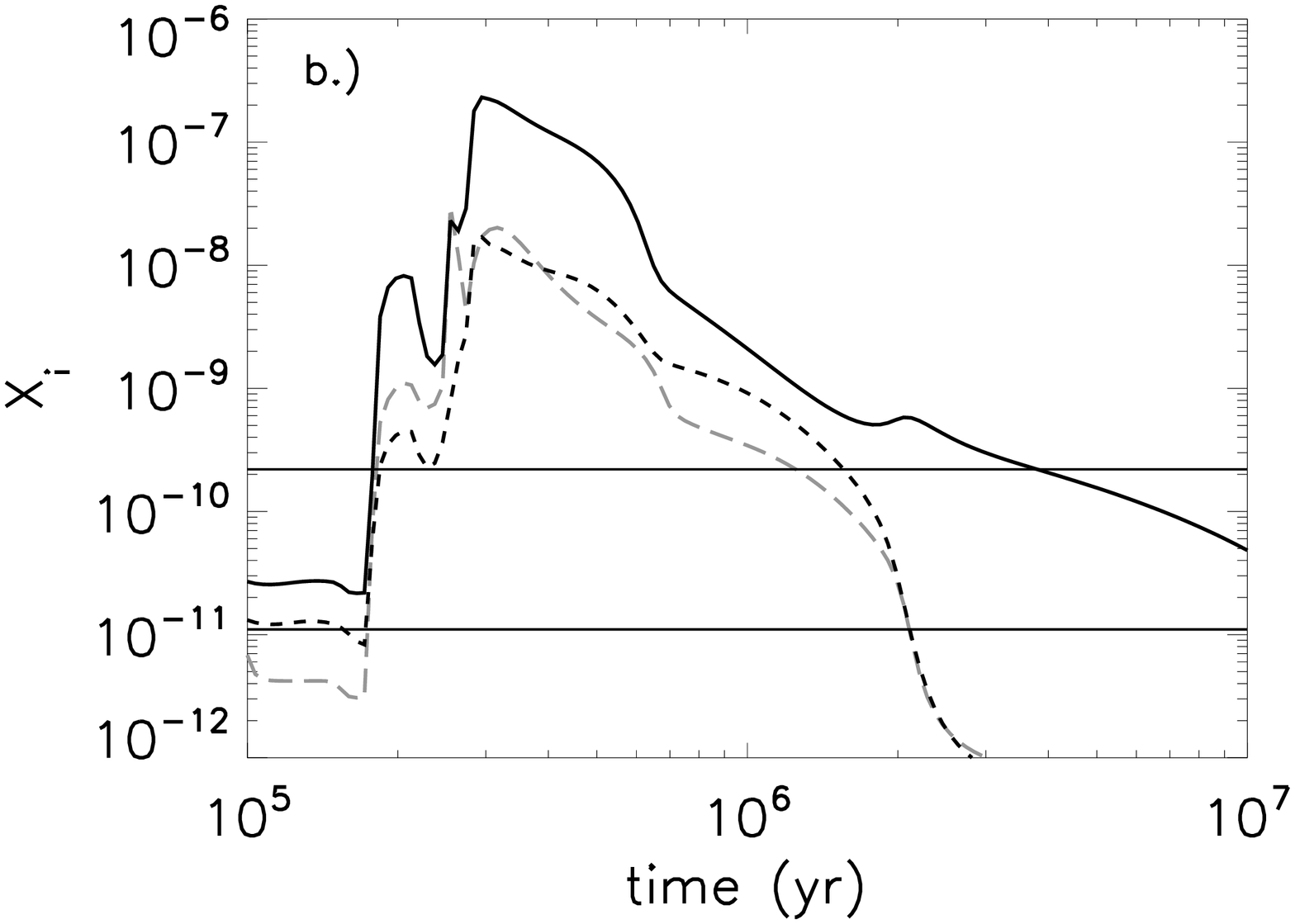}
}
\subfigure{
\includegraphics[angle=90,scale=.25]{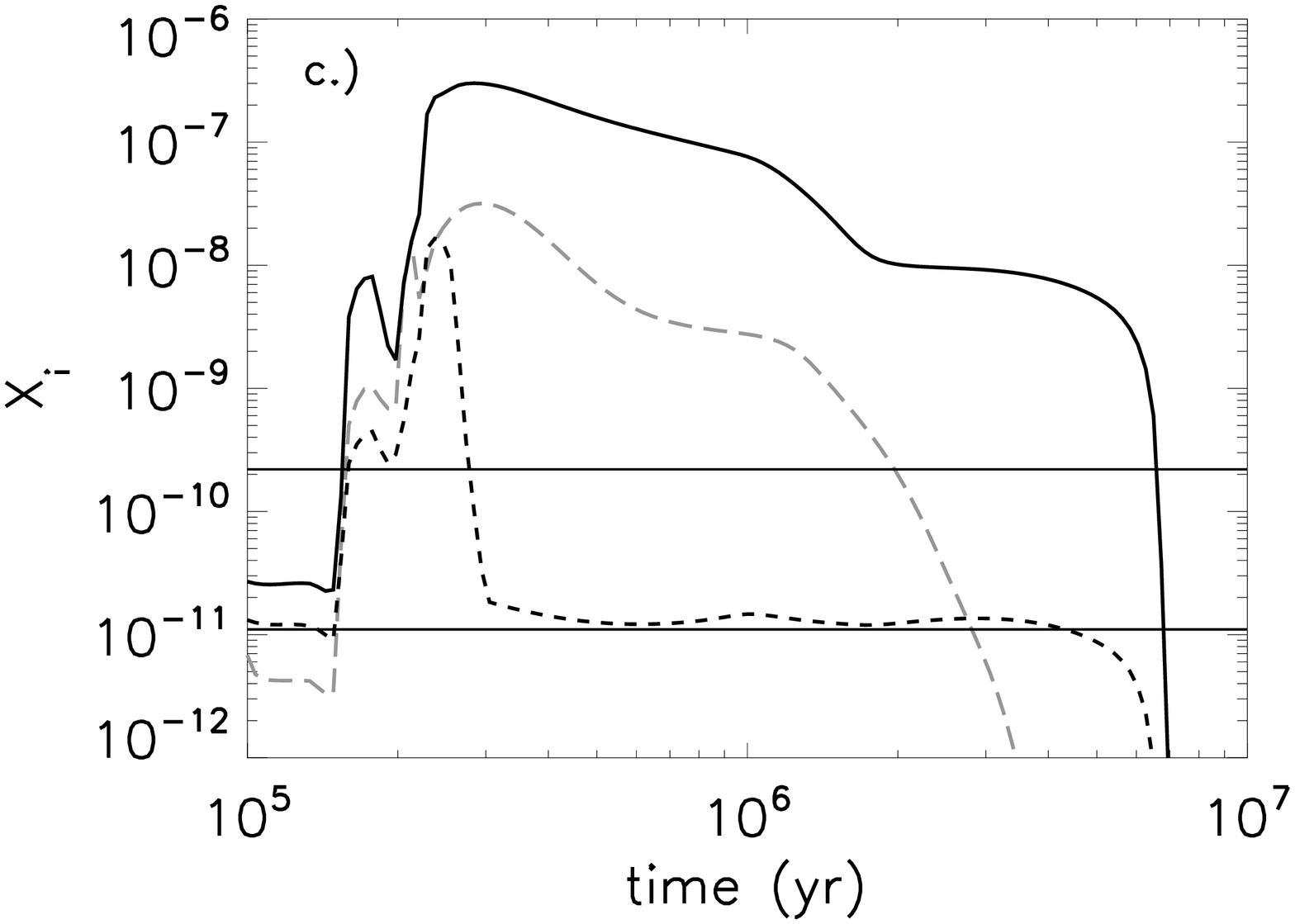}
}
\caption{Temporal evolution of the abundances of the cyclic and carbene forms of C$_3$H$_2$ for the same warm-up scenarios as in
Fig.~\ref{fig-C3H}.  The solid black lines represent the cyclic form, while the dashed black lines distinguish the carbene H$_2$C$_3$ form.  Gray long-dashed lines again represent the previous  results, in which no distinction was made between the isomers.  The horizontal lines indicate the observed abundances of the cyclic and carbene forms toward L1527 \citep{sea08,sea09b}, as labeled. \label{fig-C3H2}}
\end{figure}

We show the calculated evolution in Figure~\ref{fig-C3H2}  for c-C$_3$H$_2$ and H$_2$C$_3$ for various warm-up models in the newer network and for the previously undistinguished species C$_{3}$H$_{2}$ from HHG.  The results are more complex than expected.  The more stable c-C$_{3}$H$_{2}$ isomer is  more abundant than the carbene isomer at most times in the assorted warm-up models, although often by less than the factor of 50 assumed in HHG.  Unexpectedly, the cyclic isomer can be more abundant than the undifferentiated species in the HHG model, which is found to be a secondary effect caused by a number of new barrierless reactions in the updated network.  In addition, at late times in the $T_{\rm max}=30~$K  model an inversion of the H$_2$C$_3$ and c-C$_3$H$_2$ abundances is apparent, an effect also caused by new reactions other than those with barriers.  

The initial evolution during the warm-up progresses in a similar manner in panels b) and c) for the $T_{\rm max}=100~$K and 200 K models, but an increase in the peak abundances of the species becomes apparent as a result of desorption from the grain surfaces.  Differences in surface rates prior to the onset of desoprtion result in more C$_3$H$_2$(s) than H$_2$C$_3$(s), and ultimately affect the gas phase abundances during desorption.  The difference in the abundance of C$_3$H$_2$ between current and prior results can be attributed to the onset of C$_3$H + H$_2$ $\rightarrow$ C$_3$H$_2$ + H for $T\gtrsim 80$~K.  Panel c) illustrates a further difference between C$_3$H$_2$ and H$_2$C$_3$ in the current model as the destruction rate H$_2$C$_3$ + H$_2$ $\rightarrow$ C$_3$H$_3$ + H becomes a dominant process as $T\gtrsim 140$~K.  The C$_3$H$_3$ then reacts to preferentially form C$_3$H$_2$ rather than H$_2$C$_3$.  The energy barrier for this rate of $E_{\rm A}=1500~$K is significantly smaller than the value of 27,100 K for the corresponding destruction reaction involving C$_3$H$_2$.  This difference has the overall effect of converting not only C$_3$H but also H$_2$C$_3$ into C$_3$H$_2$ at high temperatures.      

The species C$_4$H$_2$ and C$_6$H$_2$ also exist in different isomeric forms, but these were not differentiated in the updated reaction set.  These species differ from C$_3$H$_2$ because only the carbene form has been observed.   A comparison of model results shows little discernible difference between the current results and previous HHG models for C$_4$H$_2$ at lower temperatures, but some difference at late times in the $T_{\rm max}=200~$K model.  The new C$_6$H$_2$ abundances show a more distinct enhancement over the previous ones during warm-up as $T\gtrsim 80$~K.  This suggests that the updated reaction set drives the longer-chain C$_{\rm n}$H material toward  C$_{\rm n}$H$_2$ species at high temperature conditions.   Given the comparison of  undifferentiated model results and observations for the H$_{2}$C$_{4}$ and H$_{2}$C$_{6}$ isomers,  it is likely that much of the  carbon-chain material is deposited into the unobservable forms HC$_{{\rm n}}$H. 

\subsection{Cyanopolyynes: {\rm HC$_{\rm n}$N}}
\label{sec-cyano}

The absence of viable destructive reactions with H$_{2}$ for cyanopolyyne species implies that these should not be depleted at high temperatures.  Instead, the reactions in Table~\ref{tbl-newrxns} suggest that radical C$_{\rm n}$N chains should be converted to HC$_{\rm n}$N species as $T\gtrsim 80$~K.  This further suggests that the C$_{\rm n}$N evolution might be similar to that of C$_{\rm n}$H, while the HC$_{\rm n}$N evolution might resemble C$_{\rm n}$H$_2$.    

\begin{figure}[b]
\centering
\subfigure{
\includegraphics[angle=90,scale=.25]{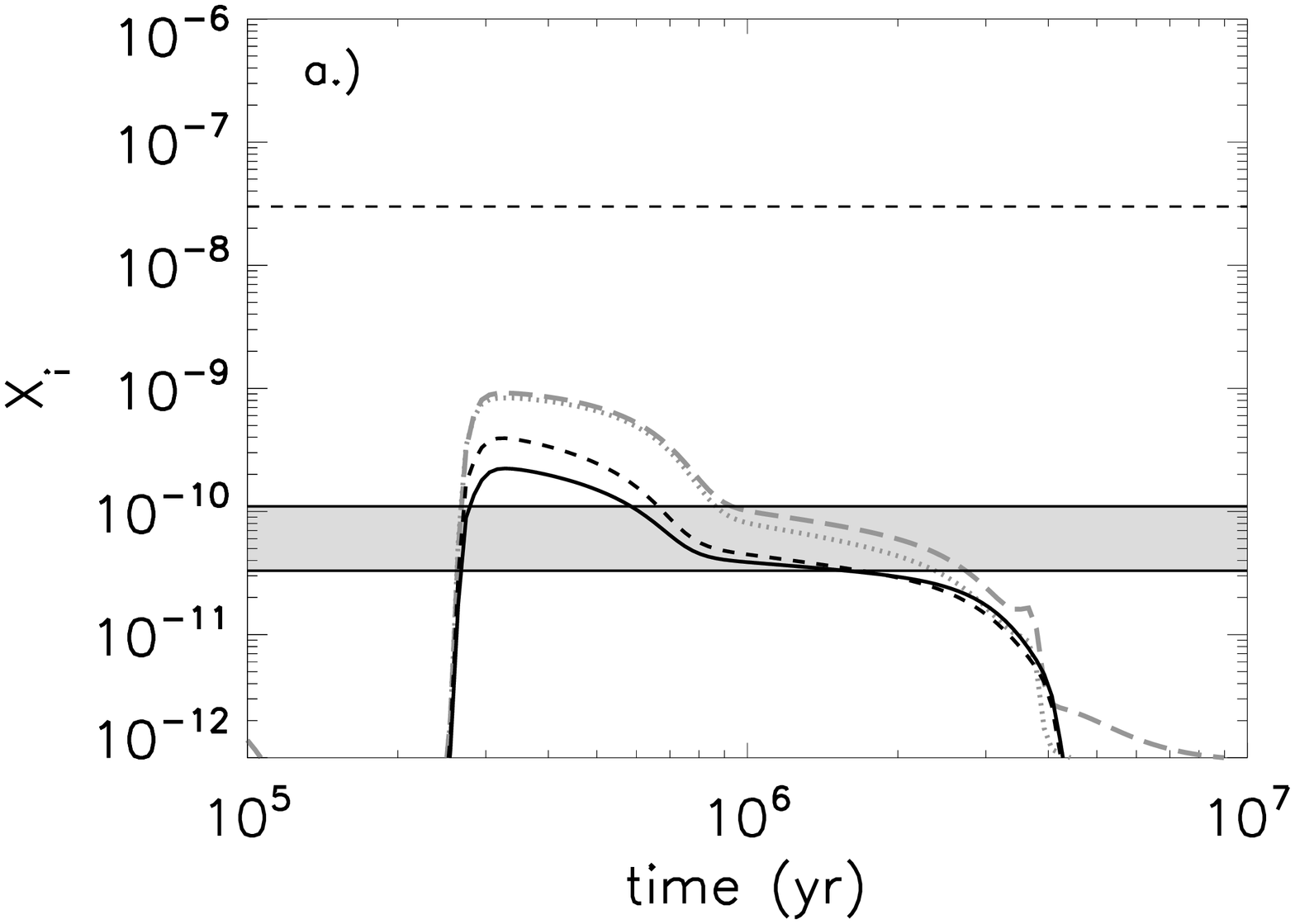}
}
\subfigure{
\includegraphics[angle=90,scale=.25]{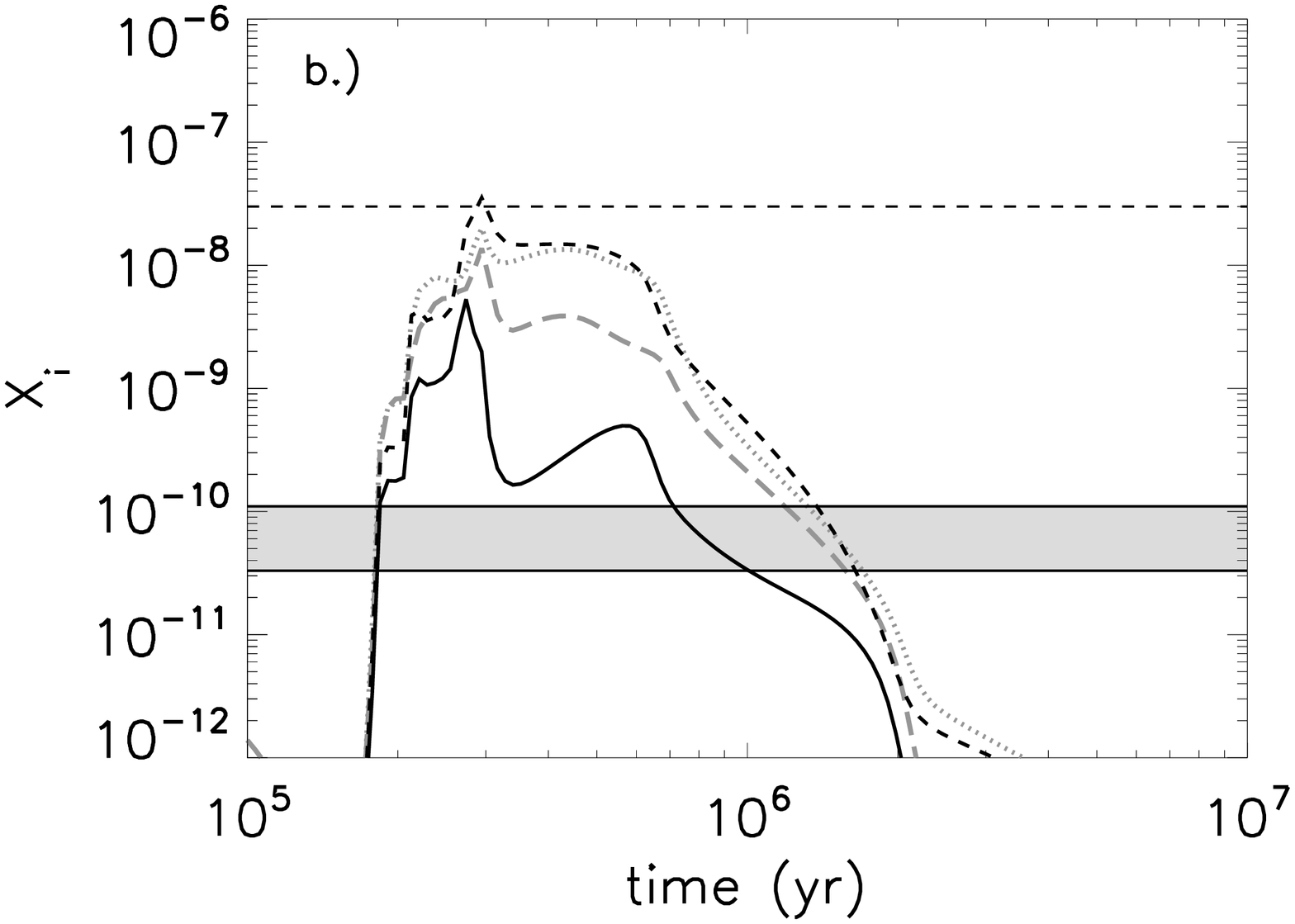}
}
\subfigure{
\includegraphics[angle=90,scale=.25]{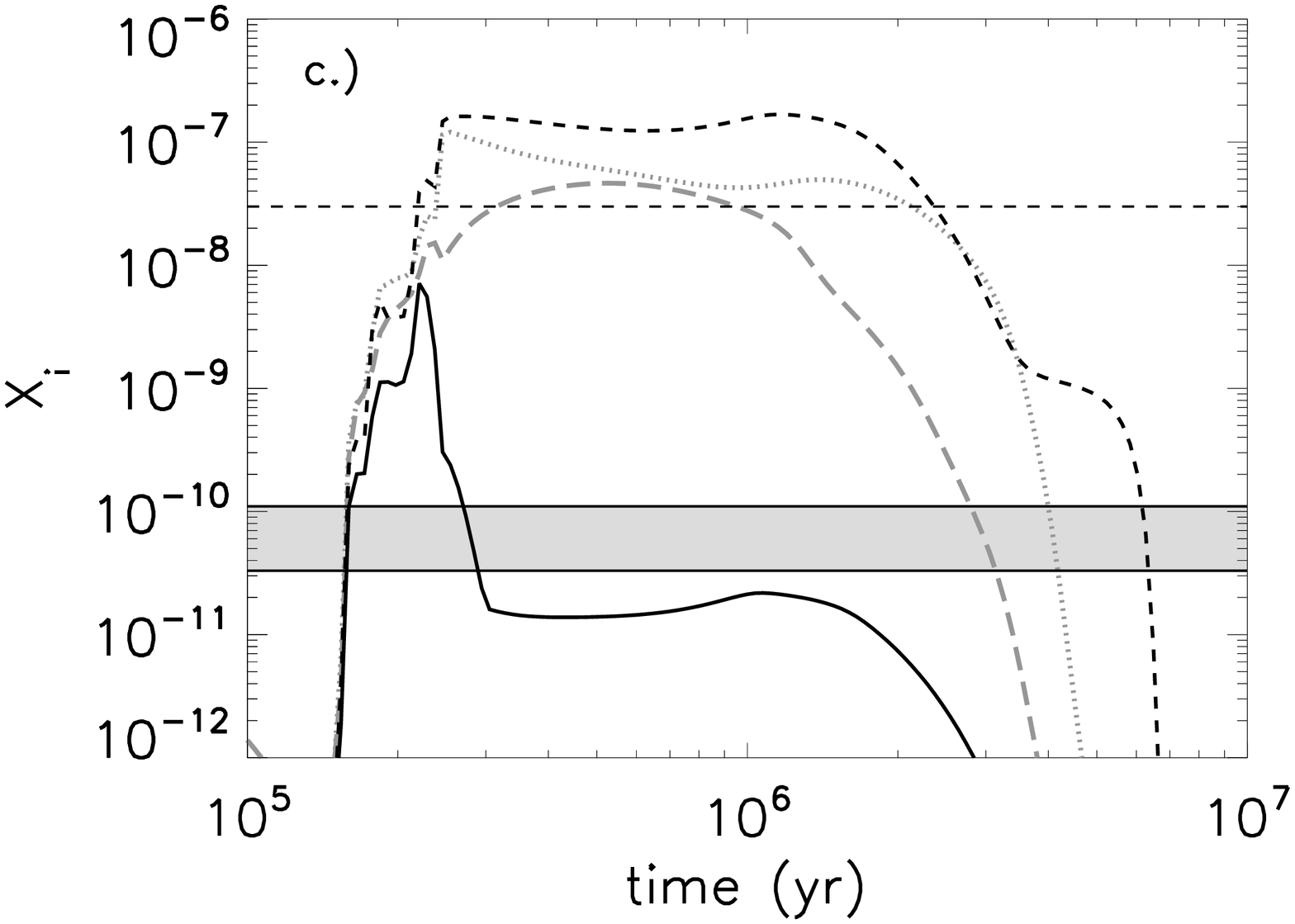}
}
\caption{Temporal evolution of the abundances of C$_5$N and HC$_5$N for the same warm-up scenarios as Fig.~\ref{fig-C3H}.  The black solid lines represent the current C$_5$N results, the black dashed lines represent the current HC$_5$N results, and the gray long-dashed and dotted lines represent the corresponding HHG results.  The horizontal gray band represents the range of observations for HC$_5$N toward lukewarm corinos, while the dashed line represents the J{\o}rgensen-Bisschop estimate of the upper limit of HC$_5$N toward hot cores.  \label{fig-cyano}}
\end{figure}

The warm-up model results for C$_5$N and HC$_5$N are illustrated in Figure~\ref{fig-cyano}, and compared with previous results from the HHG analysis.  In panel a), it can be seen that the old and new results for the $T_{\rm max}=30$~K models differ for both species, and to a larger degree than for C$_4$H.  The higher temperature results in panel c) do support the conjecture that C$_5$N will peak, and then rapidly decline with increasing $T$ in favor of HC$_5$N, in stark contrast with previous results.   This trend appears also in panel b), but to a lesser extent.        

\subsection{Methanol: {\rm CH$_3$OH}}

In contrast with the unsaturated carbon chain species, CH$_3$OH is more typically associated with hot core/corino environments \citep{lovas82,b87}.  Thus, the abundance is anticipated to increase with increasing temperature, and should not be readily destroyed by the inclusion of the high temperature network.  We include a comparison between the old and new results for the $T_{\rm max}=200$~K models in Figure~\ref{fig-4spec}d.  Some difference in abundance can be seen between the models for the range $T\approx 20-80$~K, which can be attributed to the inclusion of a photodesorption process in the updated network \citep{obergch3oh}.  As the temperature rises beyond this range, the abundance of CH$_3$OH rises by nearly four orders of magnitude as a result of thermal desorption from grain surfaces.  Little discernible difference is noted between the two models until later times following the warmup.
 
\subsection{Bare Radical Carbon Chains: {\rm C$_{\rm n}$}}
\begin{figure}[b]
\centering
\subfigure{
\includegraphics[angle=90,scale=.3]{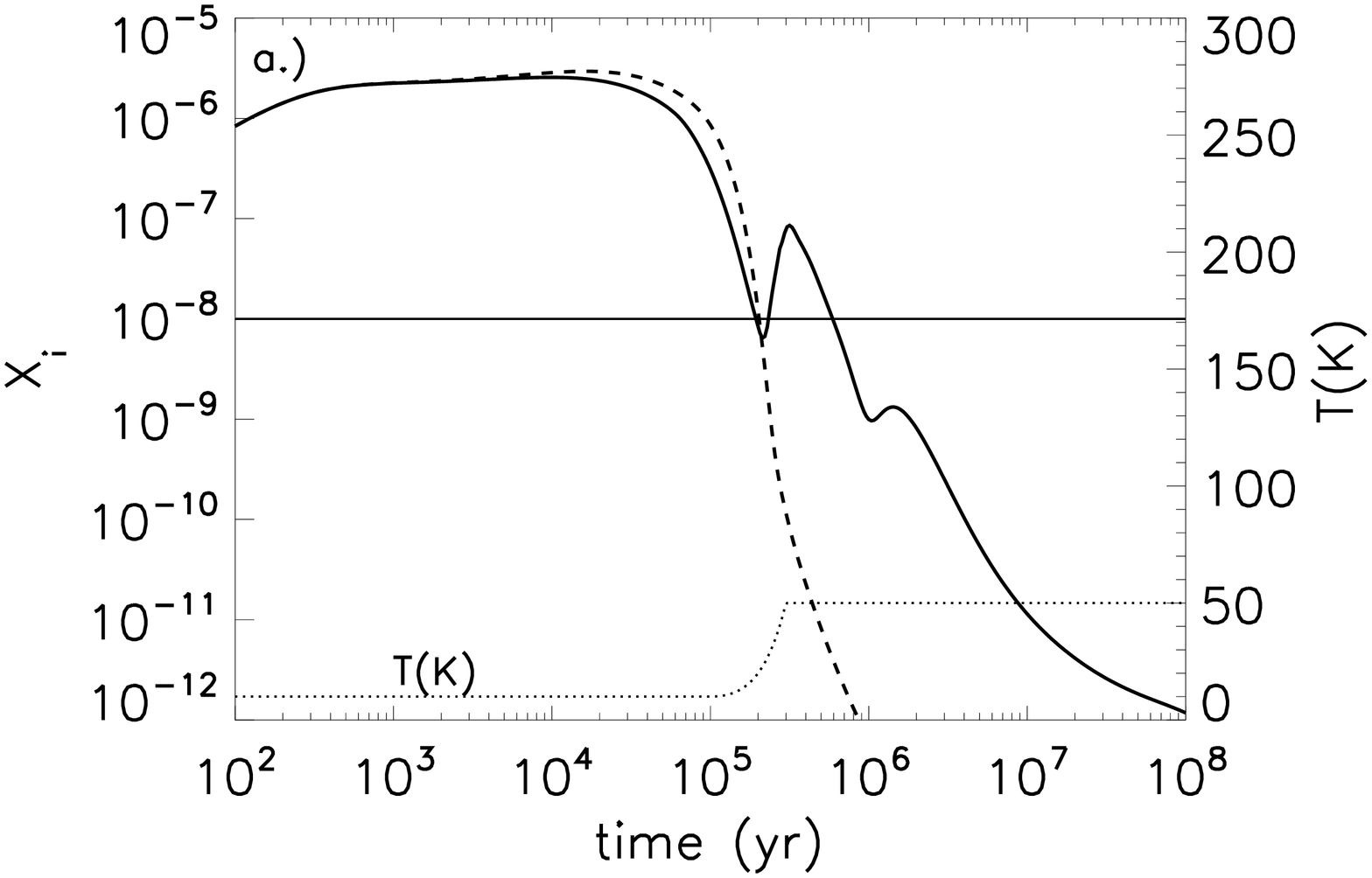}
}
\subfigure{
\includegraphics[angle=90,scale=.3]{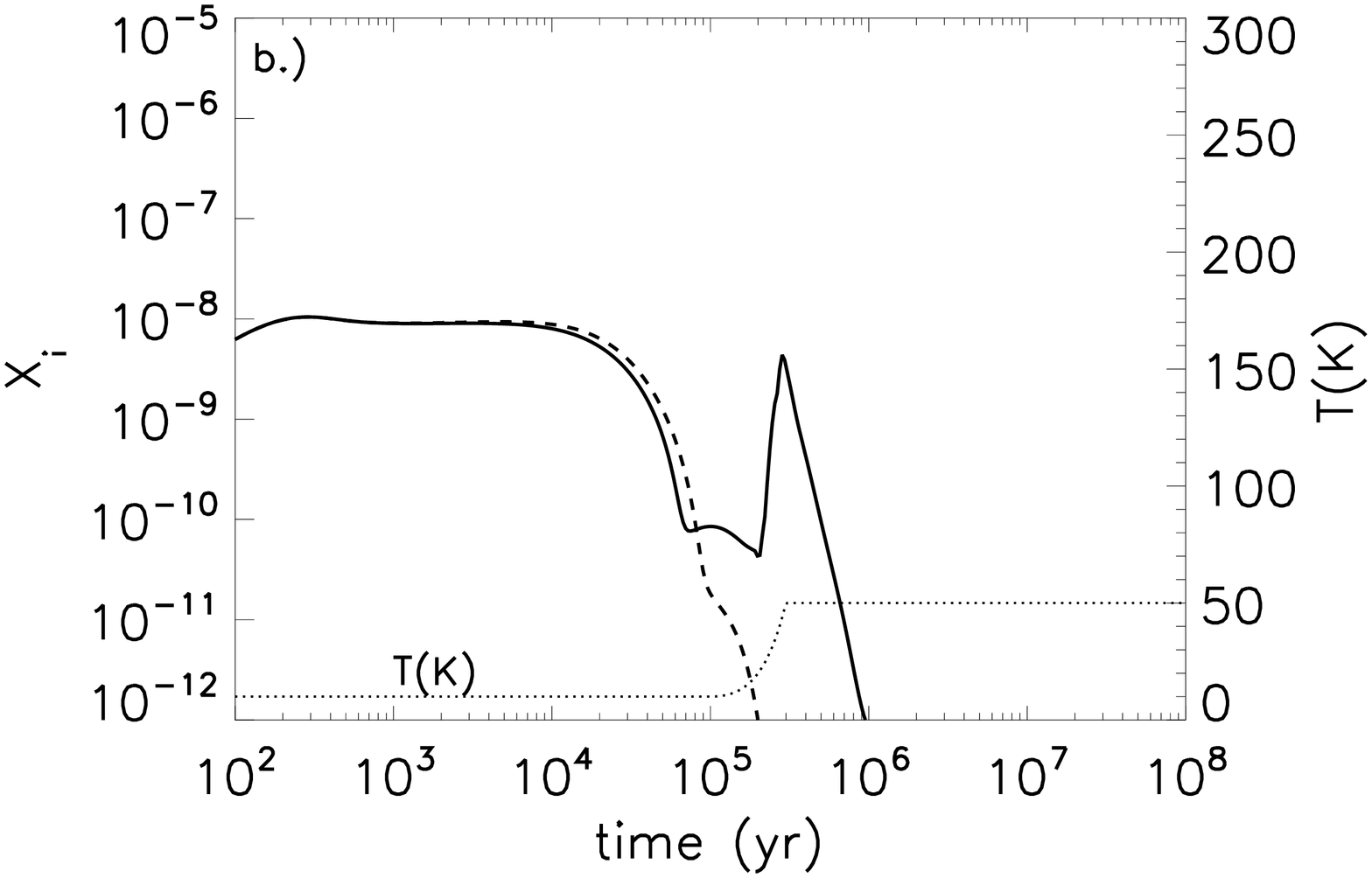}
}
\subfigure{
\includegraphics[angle=90,scale=.3]{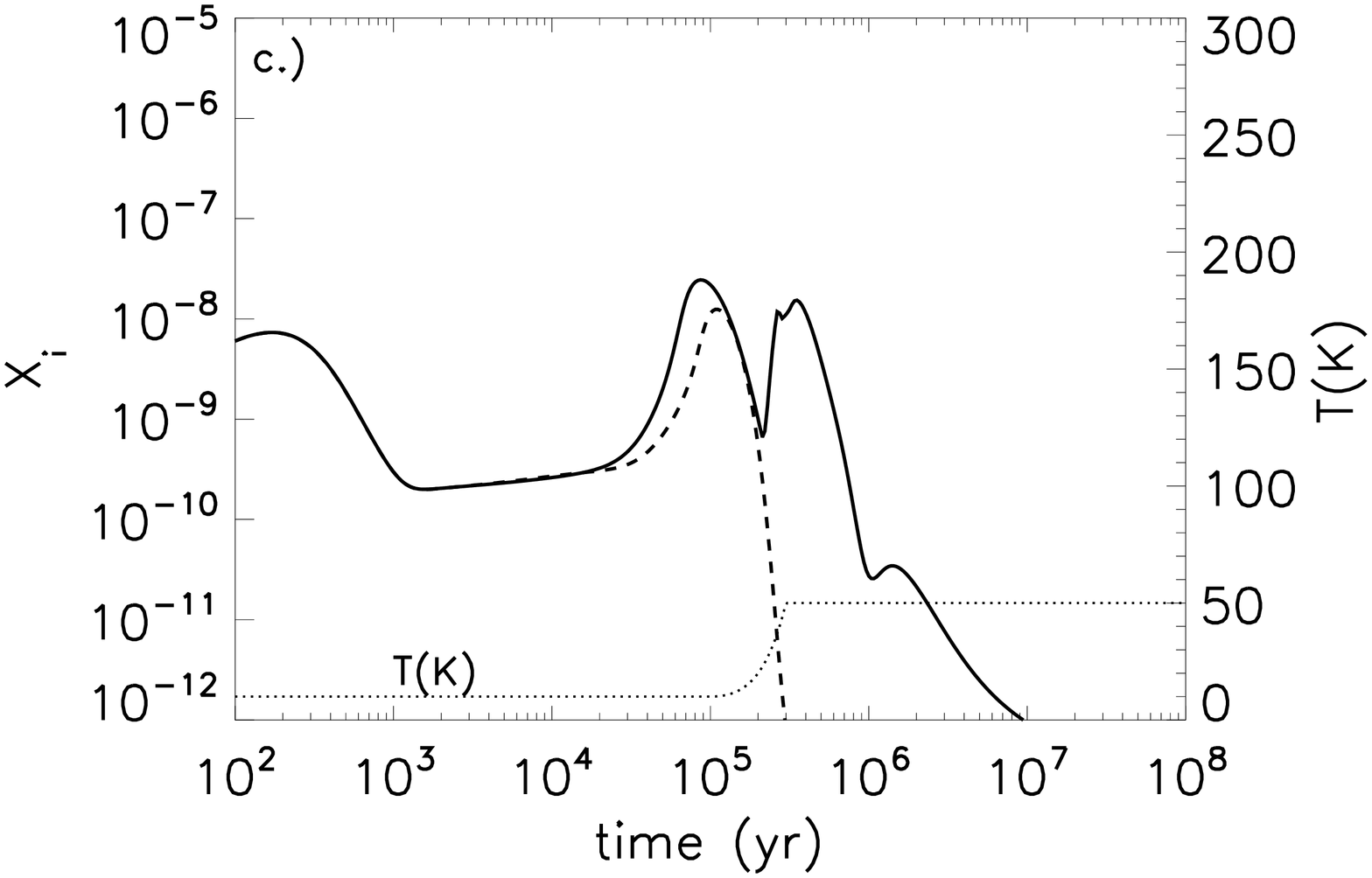}
}
\subfigure{
\includegraphics[angle=90,scale=.3]{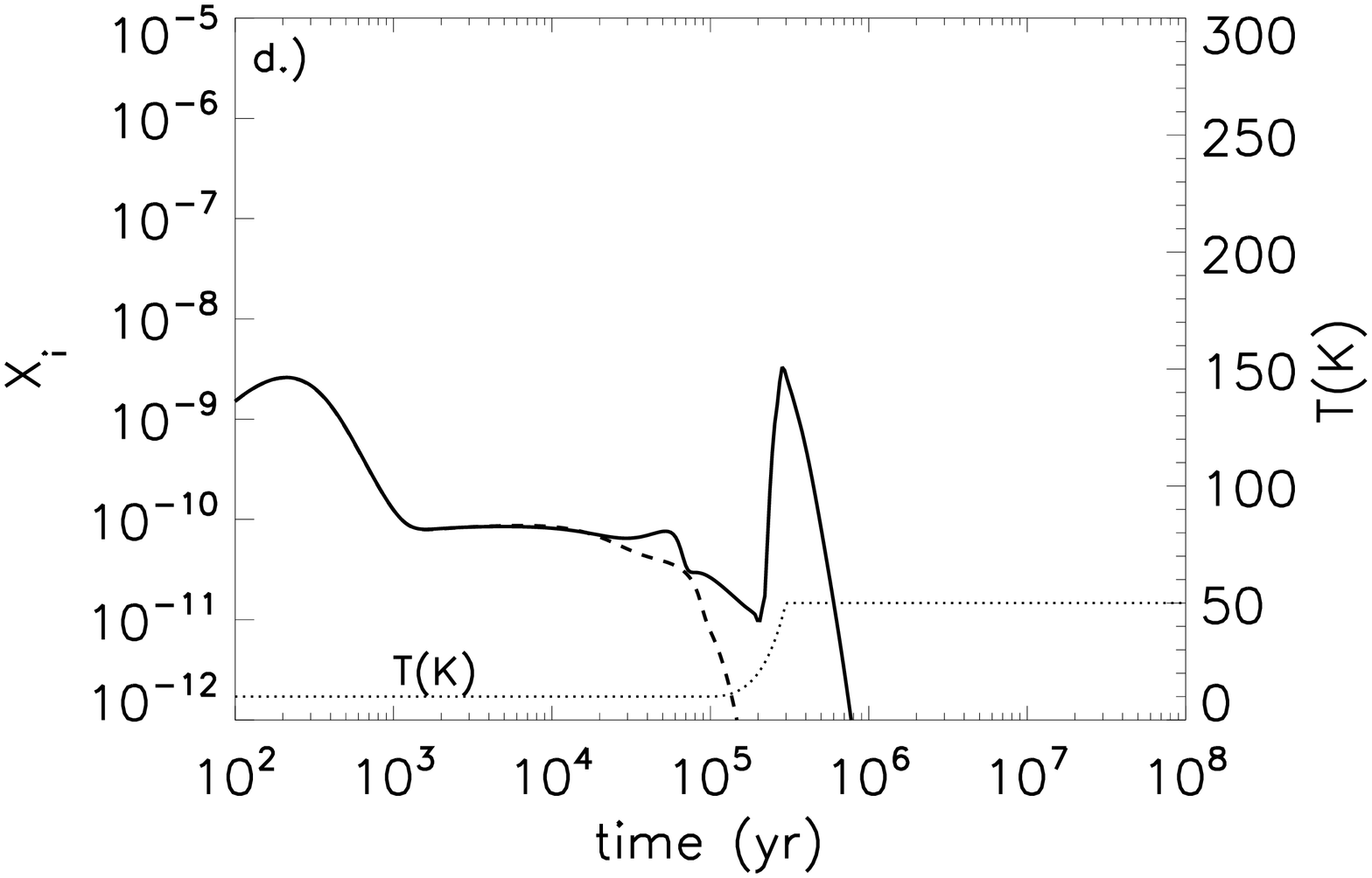}
}
\caption{Temporal evolution of the fractional abundances of four bare carbon chain species: a.) C$_3$; b.) C$_4$; c.) C$_5$; d.) C$_6$ during the warm-up period to $T_{\rm max} = 50$~K.  The solid lines represent new gas-grain results including non-zero barrier reactions, while the dashed lines represent models including only gas-phase chemistry.  The temperature profile for the warm-up is also plotted with respect to time as labeled.  The observed abundance of C$_3$ toward the massive star forming regions W31C and W49N is indicated with a horizontal line in the first panel \citep{mea10c3}. \label{fig-c3bare}}
\end{figure}

Our previous analysis included but did not focus on bare carbon chains, C$_{\rm n}$, primarily because detections of such species had not been made toward L1527.  Since then, \citet{mea10c3} have observed vibrational-rotational transitions of C$_3$ in absorption against hot-core type regions W31C and W49N using the HIFI detector on Herschel.   The absorption is thought to arise from a cooler envelope surrounding the hot cores at an estimated temperature of $\approx 50$~K.  Their results followed an earlier, more tentative interstellar detection towards Sgr B2 at low resolution.  While no larger bare carbon chains have been detected by Herschel, we have re-computed results to $n=10$ and show results between $n=3-6$ in Figure~\ref{fig-c3bare} for a model at a density  $n_{\rm H}=2 \times 10^{5}$~cm$^{-3}$ and a warm-up to $T_{\rm max}=50$~K based on the estimates of \citet{mea10c3}.  There is not a significant difference between the C$_{\rm n}$ abundances from the new models and our prior results because the $T_{\rm max} = 50$~K is not hot enough to make the C$_n$ + H$_2$ $\rightarrow$ C$_n$H + H reactions relevant as a destruction mechanism.  Therefore, we switch focus and consider the role of surface chemistry in their formation.

 For each bare carbon chain, we plot the new results in Figure~\ref{fig-c3bare} including surface chemistry and results for an analogous model without surface chemistry.  We note that with the gas-grain models, the pre-warm-up abundances of the odd-numbered chains C$_{3}$ and C$_{5}$ exceed the secondary peaks caused by the warm-up, which starts at  $t=10^5$~yr.  In the purely gas-phase models, on the other hand, there is no warm-up peak at all; rather, there is a precipitous drop in the abundances of the bare carbon chains.   These results illustrate  that the warm-up peaks clearly require surface chemistry.  However, as a more detailed analysis shows, the surface chemistry is not direct, but serves mainly to produce methane during the cold era, which evaporates as the temperature nears 30 K, and acts as a precursor for another round of ion-molecule chemistry leading to carbon chains \citep{sea07}.   Like the other carbon-chain radicals, the primary re-generation mechanism for the C$_3$ warm-up peak is directly related to acetylene derived from methane, and occurs via the gas-phase reaction C+C$_2$H$_2$ $\rightarrow$ C$_3$ + H$_2$.  During the warm-up to 50 K, all four chains reach an abundance peak of $X \sim 10^{-8}$ for approximately the same time frame of several times $10^{5}$~yr, with only a gradual decline in the peak height as the chain length increases.

\section{Comparison with Observations}
\label{sec-obs}
The motivation for  including high-temperature reactions stems partially from an increase in relevant observational data.  This increase includes new detections toward L1527, hot corinos, and HIFI sources, as well as a new candidate lukewarm corino.  The expanded data set allows for the compositional comparison of these sources.   

\subsection{Lukewarm Corinos}

Observed fractional abundances with respect to $n_{\rm H}$ for the lukewarm corino sources  L1527 and B228 are listed in Table \ref{tbl-LC}.  In addition to the new observations of B228, this table includes some new and revisited observations toward L1527 \citep{hoy09,sea09,sea09b}.  Some updates include the distinction between the c-C$_3$H$_2$ and H$_{2}$C$_{3}$ isomers, and the detections of l-C$_3$H and CH$_3$OH.  To convert observations from column densities to fractional abundances we used  a total column density of $N_{\rm H}=6\times 10^{22}$ cm$^{-2}$ for L1527 \citep{jsv02}.   Based on dust continuum flux measurements, \citet{sea09} adopted the same value for B228.  

The mean confidence level, $\kappa$, was used to quantify the agreement between model results and observed abundances based on the approach of \citet{gwh07}.  With their definition, an average confidence level of unity refers to perfect agreement, an average confidence level of 0.317 refers to an average deviation between observed and calculated abundances of an order of magnitude, while an average confidence level of 0.046 refers to an average deviation of two orders of magnitude.  This mean confidence level is calculated as a function of time for our models; the best value is referred to as $\kappa_{\rm max}$, and it occurs at a time $t_{\rm opt}$ and corresponding temperature $T$.  We also determine the number of species that agree within an order of magnitude at that time.    We used our 10 K constant-temperature model and three warm-up models ($T_{\rm max}=30$~K, 100 K, and 200 K)  to study L1527.  The results appear in Table \ref{tbl-LC}.

The inclusion of new reactions and new detections into the comparison has somewhat decreased the overall level of agreement from our previous work.  Part of the problem is the need to fit the abundances of the carbene structures without the simple approximation that they have 2\% of the polyacetylene abundances.  First, we find that the $t_{\rm opt}$ values, and thus the associated temperatures, do not change for the three warm-up models when compared with previous results.  The values of $\kappa_{\rm max}$ for the $T_{\rm max}=30$~K, 100 K, and 200 K models decrease slightly in the new analysis to 0.599, 0.580, and 0.577, corresponding to overall agreement within a factor of 3.5-4 at optimum temperatures of 25 K, 26 K, and 27 K, respectively.  In terms of number of molecules, the warm up models fit 19, 20, and 19 species out of 26 to within an order of magnitude.  This level of agreement is not much better than the agreement with the 10 K constant-temperature model of $\kappa_{\rm max}=0.514$, which occurs at a time of $4.2 \times 10^{4}$ yr, slightly before warm-up starts.  Here 19 species are fit to within an order of magnitude.
This model, however,  still significantly under-produces some major characteristic carbon chain species:  C$_2$H, C$_4$H, HC$_5$N, HC$_7$N, and HC$_9$N at the $t_{\rm opt}$ value, and must therefore be regarded as somewhat inferior for these species, whatever its reliability for non-carbon-chain species.

The newly observed source B228 is presently limited to seven species, each of which  agrees in abundance with the corresponding L1527 values within a factor of 5 or better.  The new warm-up models fit five of these, but over-produce H$_2$C$_3$ and H$_2$C$_4$ by more than an order of magnitude.  The constant temperature model is mostly consistent with the L1527 comparison, but it slightly under-produces CH$_3$CCH toward B228 while remaining at the borderline for agreement with L1527.  In summary, the updates of new species and reactions have somewhat altered the model results for L1527, but have not significantly changed the overall conclusion that the fits of warm-up models to L1527 observations near $T\sim 25$~K are a preferable explanation for the observed composition over a constant temperature model at $T = 10$~K.  The recent interferometric observations of L1527 conducted by \citet{sea11} show that the distributions of C$_{2}$H and C$_{4}$H have extended structures as well as an enhanced component, which indicate both a regenerative component, fit by our warm-up models, and a starless-core phase, corresponding to remnant cold species.  

Based on the fact that the fits to the abundance data L1527 and B228 do not change much when the maximum temperature is raised from 30 K to 200 K,  it is unclear whether lukewarm corinos represent a terminal evolutionary stage at 25 - 30 K or an intermediate step of an unfinished process leading to hot corinos, or some other stage.  It is clear from the recent interferometric study of \citet{sea11} that there are depletions of C$_2$H, C$_4$H, and c-C$_3$H$_2$ toward the innermost regions of L1527, providing some initial evidence for a third chemically distinct region, possibly a hot corino.

\subsection{Hot Corinos}

In Table~\ref{tbl-obshc}, we compile observed abundances toward the hot corinos IRAS 16293-2422, NGC 1333 IRAS 4A and 4B \citep{sea09,ssy06,bclwcccmpt04,jsv04,mea04,cea03}.   It can be seen that the first of the three is chemically richer than the other two, although the three share several chemical characteristics:  for example, both IRAS 16293-2422 and NGC 1333 IRAS 4B have measurable abundances of C$_{4}$H, which are however two orders of magnitude below that  of the lukewarm corinos \citep{sea09,cea03}.  On the other hand, C$_{2}$H and CH$_{3}$CCH have only been detected in IRAS 16293-2422; while the former has a much lower abundance than seen in the lukewarm corinos, the more saturated CH$_{3}$CCH has a considerably higher abundance \citep{sea09,cea03}. In addition, all three contain high abundances of oxygen-containing organic species, such as methanol, and methyl formate.

In Table~\ref{tbl-obshc}, we also present the results of the $T_{\rm max} = 100$~and 200 K models at $T=50~$, 100, 150 and 200 K to examine the effect of temperature on the evolution of the three carbon-chain species and methanol, where we use the results of both models to report a range of abundance at 50 and 100 K, and report only the results from the 200 K model for $T=150$ and 200 K.  These results can be compared with those for warm-up models of lukewarm corinos, shown in Table~\ref{tbl-LC}, where the best results all hover in the temperature range $25-27$~K.  Unlike the case of lukewarm corinos, we are not interested in determining the best models overall for hot corinos since our gas-grain network does not contain enough chemistry to simulate the abundances of the complex oxygen-containing organic molecules found in these sources.   This chemistry is studied in the papers of \citet{gwh08} and \citet{laas11}.

The lukewarm/hot corino distinction is also explored in the solid lines (new model results) in Figure~\ref{fig-4spec}, which includes the $T_{\rm max}=200~$K models and observed abundances for methanol and the three carbon-chain species C$_{2}$H, C$_{4}$H, and CH$_{3}$CCH.  In comparing lukewarm and hot corinos with this warm-up model, we are tacitly assuming that lukewarm corinos are earlier and hot corinos later evolutionary stages towards low-mass star formation. As the plot illustrates, the radicals C$_2$H and C$_4$H are observed to have higher abundances in the lukewarm corinos, while the more hydrogenated CH$_3$CCH and CH$_3$OH have  lower abundances.   The revised model results reproduce this observation.  Specifically, the best fit to the abundances of the four species in L1527 occurs for $T=30-40$~K, which is slightly higher than the best fit to all species in  L1527, which occurs at 25 K as listed in Table~\ref{tbl-LC}.   On the other hand, the best fit to the four species in the hot corino IRAS 16293-2422 occurs at 150 K with some uncertainty, due to sharp temperature variations around this value for individual species.   We have no overall best fit for this source, but we can conclude from the tabulated results in Table~\ref{tbl-obshc} that  at $T=100~$K all four calculated abundances lie within one order of magnitude of their observed values and at higher temperatures and later times, three of the four are fit to this criterion.  The $T=50$~K models still over-produce the polyynes and under-produce the hydrogenated species.   

\subsection{HIFI Sources}

The HIFI detection of C$_3$ toward dense 50~K regions in front of massive star forming regions \citep{mea10c3} may sample gas similar to that in lukewarm corinos.  In Figure~\ref{fig-c3bare}, we include the observed abundance of $X\left({\rm C_3}\right) \sim 10^{-8}$ for W31C and W49N \citep{mea10c3} along with  new warm-up model results with a maximum temperature $T_{\rm max} = 50$~K for C$_{3}$ - C$_{6}$  both with and without surface chemistry.  While the model result without surface reactions  crosses the observed abundance for C$_{3}$ on a steep downward trend with increasing time as 50 K is just reached, the model including the surface chemistry agrees with the observation within a factor of 2 or better for a period of approximately $6\times 10^{5}$ yr, clearly in better agreement with observation.  The abundance vs time plots for the larger bare carbon chains show an interesting trend in which the species reach a similar peak abundance of $X\left({\rm C_n}\right) \sim 10^{-8}$ within the same approximate time period, with a slight decline in peak abundance as chain length increases.   This result suggests that bare chains up to lengths of 8 or 9 carbons should be abundant enough for detection, provided that their spectral vibrational-rotational features fall within the detectable range of an instrument.  The abundance of C$_3$ is also included in Table~\ref{tbl-LC} at the various $t_{\rm opt}$ values.  Although it has not yet been reported toward L1527 or B228 and cannot be formally considered to agree, the values corresponding to the warm-up models are consistent with $X_{\rm C_{3}}=10^{-8}$ in these models.  Additional possible tests to verify this result include searching for bare carbon chains toward lukewarm corino sources, or searching for other species in the $T\sim 50~$K regions surrounding the HIFI sources. 

\section{Discussion}

We have calculated gas-grain chemical models using an expanded reaction network that includes non-zero barrier reactions \citep{har10} in order to better understand the abundances of carbon chains and other species in star-forming regions from lukewarm to hot corinos and in dense, cool regions detected by Herschel surrounding young high-mass stellar objects.  Our models include a cold core period at 10 K, followed by a warm-up period to a maximum asymptotic temperature ranging from 30 K to 200 K.  The specific temperature of a source can either be associated with an asymptotic temperature or a temperature in the range from 10 K to the asymptotic temperature of a model.  
It is not presently known if a lukewarm corino such as L1527 ($T \approx 30$~K) represents a terminal stage of chemical evolution or a transient phase on an evolutionary track toward the hot corino stage (100-200~K), but the present models allow either scenario, because, 

generally speaking, the optimum results we get are not strongly dependent on whether a source temperature is considered to be an intermediate temperature or an asymptotic temperature.  Recent observations show that a range of physical conditions and chemical compositions can exist not only toward complex hot cores \citep{mea10c3} but also towards lukewarm corinos \citep{sea11}, and the results provide a basis for future modeling investigations.   The present models are better suited to address the more specific question of whether large abundances of carbon chain species at $T\gtrsim 100$~K exist.

Our four main conclusions are as follows:

\begin{enumerate}

\item{\it The inclusion of new reactions into our gas-grain network does not significantly affect the results and conclusions obtained from previous warm-up model fits to lukewarm corino observations.} As a benchmark calculation, we revisited the simulations of the lukewarm corino L1527 and evaluated fits  using the same criteria as \citet{hhg08}.  The times of peak agreement, $t_{\rm opt}$, and associated temperatures  (25 - 30 K) do not change as a result of using an expanded reaction network with reactions containing barriers.  The values of the optimum mean confidence level, $\kappa_{\rm max}$, are only slightly diminished by comparison with the previously reported values, partly because of the addition of more species to the observation set.   Thus, the idea  that the chemistry of a lukewarm corino involves a warm regeneration of carbon chains resulting from CH$_{4}$ desorption at 25 - 30 K \citep{sea07}, incorporated into both our old and updated models, is supported.  

\item{\it The inclusion of the new reactions tends to lower carbon-chain radical abundances at higher temperatures.}  Prior models showed significantly increased abundances over significant periods of time for several carbon-chain species including the C$_{\rm n}$H and HC$_{\rm n}$N families above the L1527 observed values as the temperature increased from lukewarm corino ($T\sim 30~$K) toward hot corino/hot core levels ($T\gtrsim 100$~K).  In the updated networks, reactions with activation energy barriers between molecular hydrogen and polyyne radicals become the dominant destruction mechanism for these species in star-forming regions as temperatures exceed $T\gtrsim 80$~K.  The inclusion of these reactions reduces the abundances of most carbon chain radical species at higher temperatures.  

\item{\it Most importantly, the new observations and models begin to chemically distinguish lukewarm and hot corinos.} 
The species C$_2$H, C$_4$H, CH$_3$CCH, and CH$_3$OH have been observed with abundance differences of approximately two orders of magnitude between these types of sources, and previous models could not account for the depletion of C$_2$H and C$_4$H at larger temperatures.  Updated results at increased temperatures of $T\gtrsim 80$~K show agreement with the diminished abundances of C$_2$H and C$_4$H observed toward hot corinos compared with lukewarm ones.  The reactions most responsible for diminishing the abundances of the carbon chain (polyyne) radicals C$_2$H and C$_4$H and distinguishing lukewarm from hot corinos in their carbon chain inventories are those with molecular hydrogen, listed in Table~\ref{tbl-newrxns}.  Even though they possess small barriers, these reactions are still competitive at temperatures above 80 K.  The model results indicate that the ``excess" polyynes are primarily converted to unobservable polyacetylene forms of C$_{\rm n}$H$_2$, although it is not currently possible to directly verify this.  Unlike the case for the unsaturated polyyne radicals, the model results, in agreement with observation, show that the  the abundance of CH$_3$CCH has  the opposite trend of lower abundances in lukewarm sources and distinctively larger abundances in hot corino sources.  The observed abundances of CH$_3$OH show a similar pattern of smaller values toward a lukewarm source and larger values toward hotter sources, which the model results also accurately replicate. The trends observed in these four species provide the framework for a molecular distinction between the lukewarm corino and hot corino classes of protostellar envelopes.  A recent observation of L1527 indicates depletions toward the innermost region, suggesting that such differentiation occurs within an individual source \citep{sea11}.  This observation may be consistent with that of \citet{Oberg10} where complex oxygen-containing molecules appear to be associated with a cold portion of the low-mass protostar B1 -b.

\item {\it The high abundance of C$_{3}$ observed in absorption in dense regions with $T\sim 50$~K towards W31C and W49N is explained by warm-up models using both our previous and updated gas-grain network. }  Further, this agreement is reached without requiring the photo-destruction of PAH molecules, another suggested source of this species \citep{mea10c3}.  The new models presented here provide evidence of only a slight diminution of abundance as a function of increasing chain length for more complex bare carbon chains.  The formation of C$_3$ from acetylene during the warm-up implies the presence of lukewarm corino-like regions within the complex sources. 

\end{enumerate}

Each of these conclusions could be further tested and improved with more observational data.  The survey of \citet{sea09} has indicated that some additional sources such as B1 and L1448N may also classify as lukewarm corinos on the basis of enhanced C$_4$H abundance, while \citet{ccbwm11} indicated that the enhanced abundance of several carbon chain species toward L1251A (IRS3) could possibly be attributed to a similar effect.  We encourage follow-up observations of both lukewarm and hot corinos to investigate other carbon-chains including CH$_3$CCH, as well as CH$_3$OH.  The depleted abundances of C$_2$H and C$_4$H toward hot corinos may make their detection more difficult, but the 
CH$_3$C$_{\rm n}$H chains (n=2, 4, 6, $\dots$) should be present in enhanced abundance toward these sources.  The model results for the HIFI sources would benefit from additional detections of longer bare carbon chains or an expanded search for other  carbon chains in the inventory  of lukewarm or hot corinos.    

The authors gratefully acknowledge contributions useful discussions of observational data with Nami Sakai, Maryvonne Gerin, and Bhaswati Mookerjea.  G.H. thanks Rose Finn for computational time.  E.H. would like to thank the National Science Foundation for supporting his research program in astrochemistry and NASA for support to study the formation of pre-planetary matter and the analysis of Herschel data.

\clearpage

\clearpage

\begin{deluxetable}{lccccccc}
\tabletypesize{\scriptsize}
\tablecaption{Rate Coefficients for Selected New Classes of Reactions \label{tbl-newrxns}}
\tablewidth{0pt}
\tablehead{& \colhead{Reaction} & & &  \colhead{$n$} & \colhead{$\alpha$ (cm$^3$ s$^{-1}$)} & \colhead{$\beta$}& \colhead{$E_{\rm A}$ (K)} }
\startdata
1.) & C$_{\rm n}$H + H$_2$ & $\rightarrow$ & C$_{\rm n}$H$_2$  + H  & 2-9 & 1.14(-11) & 0 & 950 \\
2.) & C$_{\rm n}$  + H$_2$ & $\rightarrow$ & C$_{\rm n}$H  + H  & 2-9 & 1.60(-10) &   0 & 1420 \\
3.) & C$_{\rm n}$N  + H$_2$ & $\rightarrow$ & HC$_{\rm n}$N  + H  & 3,5,7,9 & 4.04(-13) &  2.87 & 820
\enddata

\end{deluxetable}

\begin{deluxetable}{lcccccc}
\tabletypesize{\scriptsize}
\tablecaption{Molecular Fractional Abundances in Lukewarm Corinos \label{tbl-LC}}
\tablewidth{0pt}
\tablehead{ & \colhead{   Observed} &  & \colhead{Model} \\ 
\colhead{Species}  & \colhead{L1527}   & \colhead{B228} & \colhead{$T=10$ K} & \colhead{$T_{\rm max}=30$ K} & \colhead{$T_{\rm max}=100$ K} & \colhead{$T_{\rm max}=200$ K} }
\startdata
\hline
C$_2$H      & 1.2(-08)\tablenotemark{\spadesuit},\tablenotemark{a} & 5.3(-09)\tablenotemark{a} &\bf 3.0(-10) &  8.9(-09)     & 9.9(-09)     & 1.0(-08)   \\
l-C$_3$H      & 5.5(-11)\tablenotemark{b}                         & \dots                        & 2.4(-10)  &  2.3(-10) & 2.5(-10) & 2.6(-10)   \\
C$_4$H      & 3.2(-09)\tablenotemark{a}                         & 2.3(-09)\tablenotemark{a}     &\bf 1.0(-10) &  2.4(-09)     & 1.8(-09)     & 1.6(-09)  \\  
C$_5$H      & 1.6(-11)\tablenotemark{c}                         & \dots                         & 1.2(-11) &  \it 4.7(-10)\tablenotemark{\dagger} & \it 3.5(-10) & \it 3.4(-10)   \\
C$_6$H      & 1.0(-11)\tablenotemark{c}                         & \dots                         & 1.3(-11) &  \it 2.3(-10) & \it 1.4(-10) & \it 1.2(-10)   \\
c-C$_3$H$_2$& 2.2(-10)\tablenotemark{b}                         & \dots                         & 6.8(-10) &  \it 4.6(-09) & \it 3.8(-09) & \it 3.8(-09)  \\
H$_2$C$_3$& 1.0(-11)\tablenotemark{c}                         & 5.5(-12)\tablenotemark{a}     & 4.1(-11) &  \it 2.5(-10) & \it 2.4(-10) & \it 2.5(-10)\\
H$_2$C$_4$  & 2.7(-11)\tablenotemark{a}                         & 6.2(-12)\tablenotemark{a}     & 6.5(-12) &  1.6(-10)\tablenotemark{\star}     & 9.8(-11) & 8.8(-11) \\
H$_2$C$_6$  & 2.5(-12)\tablenotemark{c}                         & \dots                         & 5.2(-13) &  \it 3.4(-11)     & 1.6(-11) & 1.4(-11) \\ 
CH$_3$CCH   & 1.0(-09)\tablenotemark{a}                         & 1.7(-09)\tablenotemark{a}     & 1.2(-10) &  5.3(-09)     & 4.3(-09) & 4.3(-09) \\
CCS         & 8.5(-11)\tablenotemark{c}                         & \dots                         & 8.3(-11) &  1.1(-10)     & 8.1(-11) & 6.8(-11) \\   
HC$_5$N     & 1.1(-10)\tablenotemark{b}                         & 3.3(-11)\tablenotemark{a}     & \bf 5.0(-13) &  1.7(-10)     & 2.4(-10) & 2.4(-10) \\
HC$_7$N     & 2.7(-11)\tablenotemark{c}                         & \dots                         & \bf 4.2(-14) &  3.4(-11)     & 3.8(-11) & 3.4(-11)\\
HC$_9$N     & 2.5(-12)\tablenotemark{c}                         & \dots                         & \bf 2.4(-15) &  6.7(-12)     & 5.5(-12) & 4.5(-12)\\
HCO$_2$$^+$ & 1.0(-12)\tablenotemark{d}                         & 4.5(-12)\tablenotemark{a}     & 1.6(-12) &  8.3(-13)     & 7.1(-13) & 6.9(-13)\\
NH$_3$      & 8.3(-09)\tablenotemark{e}                         & \dots                         & 2.8(-09) &  4.4(-09)     & 6.8(-09) & 1.0(-08) \\
CO          & 3.9(-05)\tablenotemark{f}                         & \dots                         & 6.3(-06) &  3.6(-05)     & 4.3(-05) & 4.5(-05)\\
CN          & 8.0(-11)\tablenotemark{f}                         & \dots                         & 1.1(-10) &  \it 1.1(-09) & \it 2.7(-09) & \it 3.1(-09) \\
CS          & 3.3(-10)\tablenotemark{f}                         & \dots                         & 6.1(-10) &  3.3(-09)     & 2.4(-09) & 2.0(-09)\\
HCN         & 1.2(-09)\tablenotemark{f}                         & \dots                         & 3.2(-10) &  1.2(-09)     & 2.8(-09) & 3.8(-09)\\
HNC         & 3.2(-10)\tablenotemark{f}                         & \dots                         & 2.8(-10) &  1.1(-09)     & 2.7(-09) & \it 3.6(-09)\\
HC$_3$N     & 2.0-4.5(-10)\tablenotemark{b}                     & \dots                         &\bf 9.5(-12) &  5.0(-10)     & 9.6(-10) & 1.0(-09)\\   
HCO$^+$     & 6.0(-10)\tablenotemark{f}                         & \dots                         & 4.3(-10)&  1.7(-09)     & 1.7(-09) & 1.7(-09)\\
N$_2$H$^+$  & 2.5(-10)\tablenotemark{f}                         & \dots                         & 5.1(-11) &  \bf 1.1(-11)\tablenotemark{\ddagger} & \bf 9.1(-12) & \bf 8.7(-12) \\
SO          & 1.4(-10)\tablenotemark{f}                         & \dots                         & 6.6(-10) &  3.2(-10)     & 2.5(-10) & 2.1(-10) \\
CH$_3$OH    & 1.1(-09)\tablenotemark{b}                         & \dots                         &\bf 3.9(-11) &  4.2(-10)     &  1.8(-10) &  2.1(-10)\\
C$_3$       & \dots                                             & \dots                         & 6.8(-08)   & 6.4(-09)       & 6.2(-09)   & 5.2(-09)
 \\   
\hline
$t_{\rm opt}$ (yr) & \dots                                      & \dots                     & 4.2(+04) & 2.7(+05)      & 1.8(+05)     & 1.6(+05)\\
$T$ (K)            & \dots                                      & \dots                     & 10 & 25            & 26           & 27 \\
$\kappa_{\rm max}$ & \dots                                      & \dots                     & 0.514 & 0.599        & 0.580        & 0.577 \\ 
\# fits (26)       & \dots                                      & \dots                     & 19 & 19            & 20           & 19
\enddata
\tablenotetext{\spadesuit}{$a(-b) = a\times 10^{-b}$}
\tablenotetext{\star}{Results for C$_4$H$_2$ $\&$ C$_6$H$_2$ multiplied by 0.02 to approximate  abundances of observable carbene isomers}
\tablenotetext{\dagger}{Italic type indicates over-production by 1 order of magnitude or more}
\tablenotetext{\ddagger}{Bold type indicates under-production by 1 order of magnitude or more}
\tablenotetext{a}{\citet{sea09}}
\tablenotetext{b}{\citet{sea09b}}
\tablenotetext{c}{\citet{sea07}}
\tablenotetext{d}{\citet{sea08}}
\tablenotetext{e}{\citet{hoy09}}
\tablenotetext{f}{\citet{jsv04}}
\end{deluxetable}

\begin{deluxetable}{lccccccc}
\tabletypesize{\scriptsize}
\tablecaption{Molecular Fractional Abundances in Hot Corinos  \label{tbl-obshc}}
\tablewidth{0pt}
\tablehead{ & IRAS 16293-2422   & NGC 1333 IRAS 4A & NGC 1333 IRAS 4B & $T=50$~K & $T=100$~K & $T=150$~K & $T=200$~K\\
\colhead{Species} & \colhead{$X_{\rm obs}$} & \colhead{$X_{\rm obs}$}& \colhead{$X_{\rm obs}$} & \colhead{$X$} & \colhead{$X$} & \colhead{$X$} & \colhead{$X$}}
\startdata
C$_2$H        & 5.0(-11)\tablenotemark{a} & \dots  & \dots & \it 1.7-2.0(-09)\tablenotemark{\dagger}  & 2.2-5.6(-10) & \bf 4.0(-12)\tablenotemark{\ddagger} & \bf 8.5(-13)\\ 
C$_4$H        & 3.8(-12)\tablenotemark{a} & \dots &6(-11)\tablenotemark{a} & \it 2.8-3.7(-09) & 2.5-4.8(-10)     & 4.8(-12)     & 1.2(-12) \\           
CH$_3$CCH     & [2.6$\pm$0.3](-07)\tablenotemark{b} (A)  & \dots & \dots  & \bf 1.8-24.(-10) & 0.8-1.2(-06)     & 7.4(-07)     & 5.0(-07) \\
CH$_3$OH      & 3(-07)\tablenotemark{b}   & $\leq$7(-09)\tablenotemark{c}          &\dots & \bf 3.7-3.9(-10) & 1.7-8.2(-07)     & 2.3(-06)     & 1.4(-06) \\   
CO            & 3.3(-05)\tablenotemark{d} & 7.9(-06)\tablenotemark{d} &1.3(-05)\tablenotemark{d}  & 4.3-4.5(-05) & 4.5-4.6(-05)     & 4.7(-05)     & 4.8(-05) \\
CS            & 3.0(-09)\tablenotemark{d} & 1.0(-09)\tablenotemark{d} &1.2(-09)\tablenotemark{d}  & 1.2-1.4(-08) & \it 5.0-5.8(-08)& \it 5.0(-08) & \it 5.0(-08)  \\           
SO            & 4.4(-09)\tablenotemark{d} & 4.6(-09)\tablenotemark{d} &3.0(-09)\tablenotemark{d}  & 0.1-1.5(-09) & 4.1-4.7(-10)     & \bf 1.7(-10) & \bf 2.5(-10) \\
HCO$^+$       & 1.4(-09)\tablenotemark{d} & 4.3(-10)\tablenotemark{d} &6.2(-10)\tablenotemark{d}  & 1.3-2.1(-10) & 0.7-1.9(-10)     & \bf 1.6(-12) & \bf 2.0(-12) \\
N$_2$H$^+$    & 1.4(-10)\tablenotemark{d} & $>$1.0(-09)\tablenotemark{d} &3.2(-09)\tablenotemark{d} &\bf 3.8-4.7(-12) &  \bf 3.0-5.0(-12) & \bf 1.3(-13) & \bf 1.7(-13) \\
HCN           & 1.1(-09)\tablenotemark{d} & 3.6(-10)\tablenotemark{d} &2.0(-09)\tablenotemark{d} & \it 0.5-1.0(-06) & \it 1.3-2.1(-07) & \it 3.9(-07) & \it 6.1(-07) \\  
HNC           & 6.9(-11)\tablenotemark{d} & 2.8(-11)\tablenotemark{d} &1.4(-10)\tablenotemark{d} & \it 2.5-3.8(-07) & \it 0.9-1.5(-07) & \it 2.5(-07) & \it 3.1(-07)  \\  
CN            & 8.0(-11)\tablenotemark{d} & 3.7(-11)\tablenotemark{d} &1.4(-10)\tablenotemark{d} & \it 1.6-2.1(-07) & 0.3-2.7(-09) & 1.1(-10)     & 6.5(-11)\\  
CH$_3$CN      & [1.0$\pm$ 0.4](-09)\tablenotemark{b}  & [1.6$\pm$ 0.2](-09)\tablenotemark{c}  &\dots & 4.8-6.3(-09) & \it 2.1-3.6(-08) & \it 1.9(-08) & 1.3(-08) \\
CH$_3$OCH$_3$ & [2.4$\pm$ 3.7](-07)\tablenotemark{b}  & $\leq$2.8(-08)\tablenotemark{c}      &$\leq$1.2-8.0(-08)\tablenotemark{e}  &\bf 1.7-4.4(15) & \bf 0.07-6.1(-10) & \bf 1.6(-08) & \bf 1.8(-08) \\
CH$_3$CHO     & 1.9(-08)\tablenotemark{b} (A)  & \dots & \dots & \bf 1.1-2.4(-11) & \bf 0.4-1.2(-11) & \bf 8.3(-13) & \bf 7.2(-13) \\ 
              & 3.2(-08)\tablenotemark{b} (E)  & \dots & \dots & \dots & \dots           & \dots        &\dots\\ 
H$_2$CO       & 1(-07)\tablenotemark{b}   & 2(-08)\tablenotemark{f}              &\dots & 1.1-6.0(-08) & 6.8-8.7(-08)   & 2.6(-08)     & 1.1(-08) \\ 
HC$_3$N       & 1.5(-10)\tablenotemark{d} & 7.2(-11)\tablenotemark{d}  &1.1(-10)\tablenotemark{d} & \it 3.4-4.0(-08) & \it 0.4-1.2(-07) & \it 4.9(-08) & \it 2.8(-08) \\
HCOOH         & 6.2(-08)\tablenotemark{b} & [4.6$\pm$ 7.9](-09)\tablenotemark{c}  &\dots & \bf 4.2-8.7(-10) & 0.3-2.1(-08)  & 9.6(-08)     & 9.2(-08) \\ 
HCOOCH$_3$    & [1.7$\pm$ 0.7](-07) (A)\tablenotemark{b}    & [3.4$\pm$ 1.7](-08) (A)\tablenotemark{c} & 0.7-3.5(-08)\tablenotemark{e} & \bf 0.8-2.0(-12) & \bf 0.4-4.6(-10) & \bf 4.0(-09) & 3.1(-09)\\      
              & [2.3$\pm$ 0.8](-07) (E)\tablenotemark{b}    & [3.6$\pm$ 0.7](-08) (E)\tablenotemark{c} & 0.5-2.7(-08)\tablenotemark{e} & \dots          &\dots   & \dots        &\dots 
\enddata
\tablenotetext{a}{\citet{sea09}}
\tablenotetext{b}{\citet{cea03}}
\tablenotetext{c}{\citet{bclwcccmpt04}}
\tablenotetext{d}{\citet{jsv04}}
\tablenotetext{e}{\citet{ssy06}}
\tablenotetext{f}{\citet{mea04}}
\tablenotetext{\dagger}{Italic type indicates over-production by 1 order of magnitude or more}
\tablenotetext{\ddagger}{Bold type indicates under-production by 1 order of magnitude or more}
\end{deluxetable}


\end{document}